\documentclass[aps,prD,reprint,superscriptaddress,amsmath,amssymb]{revtex4-2}
\usepackage{color}
\usepackage{amsmath,amssymb}
\usepackage{graphicx}
\usepackage[colorlinks, linkcolor=blue, citecolor=blue, urlcolor=blue, breaklinks=true]{hyperref}
\usepackage{subfig}
\usepackage{mathtools}

\begin{document}
\title{Orbital angular momentum spectrum and entanglement in a rotating accelerated reference frame}
\author{Haorong Wu}
\affiliation{Department of Physics, Xiamen University, Xiamen 361005, China}
\author{Xilong Fan}
\email{xilong.fan@whu.edu.cn}
\affiliation{School of Physics and Technology, Wuhan University, Wuhan 430072, China}
\author{Lixiang Chen}
\email{chenlx@xmu.edu.cn}
\affiliation{Department of Physics, Xiamen University, Xiamen 361005, China}

\begin{abstract}
The definition of a particle varies across different theories. The quantum field theory in curved spacetime shows that from the perspective of a linearly accelerated observer, an inertial empty space may be full of thermal particles. This effect is known as the Unruh effect. When the degrees of freedom of orbital angular momentum (OAM) are considered, all OAM modes share the same expected particle number. Here, we examine the OAM spectrum in a rotating accelerated reference frame to see how the spectrum differs from the linear accelerated case. When the observer starts to rotate, not all OAM modes are allowed and some negative energy modes show up. To understand how a rotating accelerated observer actually perceives these particles, the Unruh-DeWitt detector and its detailed balance are studied. This relation is studied both in the comoving inertial frame and in the rest frame. Based on these results, the OAM entanglement degradation is explored in two-dimensional and high-dimensional cases, respectively. The results indicate that the entanglement dimension and the highest order of OAM modes are mainly related to the acceleration and the rotation, respectively. It is then demonstrated that these results can be generalized to all stationary trajectories.
\end{abstract}
\maketitle

\section{Introduction}
The definition of particle varies across different theories. A physicist before the 20th century would likely have told you that a particle is a localized pointlike object. It has some intrinsic properties that are related to some fundamental interactions. In quantum mechanics and quantum field theories, the concept of particles arises when we quantize a field \cite{peskin2018introduction}. Take the scalar field, for example. One may first identify the field variable $\psi$ and its conjugate momentum $\Pi$. Then, their quantum-mechanical commutation relations can be deduced from their classical Poisson bracket relations. By expanding the field with its eigenfunctions, one can further identify its creation and annihilation operators, i.e., $a$ and $a^\dagger$, and their commutation relations. We now can create a particle from the vacuum by using a creation operator $a^\dagger$ on the vacuum state, or annihilate one by using an annihilation operator $a$.

The story of particles becomes more complicated when the theories of relativity show up. The structure of spacetime induces several problems. First, the positivity of the norm is now questionable. The inner product should satisfy the requirements of symmetry, linearity, and observables being Hermitian. Thus, for relativistic particles, the inner product should be generalized to \cite{carroll2019spacetime}\begin{equation}
\left <g_{I},h_{J}\right >=i\int_\Sigma \big [g_{I}^*\partial_0 h_{J} - h_{J} \partial_0g_{I}^*\big ]n^\mu \sqrt \gamma d^3x, \label{eq.innerProduct}
\end{equation}
where $g_{I}=g_{I}(x^\mu)$, $h_{J}=h_{J}(x^\mu)$, $I$ and $J$ represent a set of possible indices, $\Sigma$ is a spacelike hypersurface on which the integration is carried out, $\gamma_{\mu\nu}$ is the induced 3-metric on this hypersurface, $\gamma=\det( \gamma_{\mu\nu})$, and $n^\mu$ is the unit normal vector. We will come back to this point later. If one uses this inner product to calculate the norm of an eigenfunction, (s)he may find that the result could be positive, negative, or even zero. This problem can be solved by choosing positive-norm states to associate with an annihilation operator $a_I$; then those negative-norm states will associate with a creation operator $a^\dagger_I$, so the negative-norm states are just complex conjugates of positive-norm states. Also, zero-norm eigenstates can be treated as linear combinations of positive-norm and negative-norm states. The second problem is the choice of reference frames. For nonrelativistic particles, its quantization is carried in a fixed spacetime background, and the particle states in different reference frames are unitary equivalent. However, in the theories of relativity, different reference frames may not share equivalent particle definitions. To be concrete, the Unruh effect states that a vacuum space seen by an inertial observer may be drowned by a thermal particle bath when observed by an accelerated observer \cite{unruh1976notes,parikh2000hawking,visser2003essential,russo1992end,page2005hawking,robinson2005relationship,carroll2019spacetime,martin2010population,martin2010unveiling,alsing2003teleportation,fuentes2005alice,alsing2006entanglement,adesso2007continuous,ling2007quantum,takagi1986vacuum,wu2023orbital}. Moreover, a recent study finds that if an observer has a rotational vortex structure in the transverse dimensions and carries a well-defined orbital angular momentum (OAM), it can absorb or emit Rindler particles with the same OAM when interacting with the background thermal bath \cite{bu2021unruh}. There is another problem: Will conserved quantities which are used to describe particles still be conserved in any frames? Especially, we want the particle number to be constant so that when an observer measures particles, there will be no particles coming out of nowhere or suddenly disappearing. This requires the Hamiltonian, defined in the observer's rest frame, to be time-independent, i.e., stationary. Otherwise, Hamiltonians at different times may not commute with each other, yielding inconsistent particle numbers. This further requires that the metric experienced by the observer be stationary, as well. Thus, we know that the tangent vectors of the trajectory followed by the observer must form a Killing vector field. This kind of trajectory is called a stationary trajectory. In the Minkowski spacetime, the generators of Killing vector fields are time and space translation generators ($\partial_t,~\partial_x,~\partial_y,~\partial_z$), rotation generators ($x\partial_y-y\partial_x,~y\partial_z-z\partial_y,~z\partial_x-x\partial_z$), and boost generators ($x\partial_t+t\partial_x,~y\partial_t+t\partial_y,~z\partial_t+t\partial_z$). Stationary trajectories formed by these generators can fall into six classes, from class A to class F \cite{letaw1982stationary}. The six classes can be divided into two different categories, based on how they define the vacuum. Classes A, C, and D define the Minkowski vacuum, while the other three classes define the so-called Fulling-Unruh vacuum \cite{letaw1980quantized,letaw1981quantized,letaw1981stationary,letaw1982stationary,korsbakken2004fulling}. These two vacua are not equivalent, as will be shown later.

In our previous study \cite{wu2023orbital}, we found that for a linearly accelerated observer, the OAM spectrum of the Rindler particles is uniform, because the spacetime on the transverse plane is isotropic and circularly symmetric. Here, we consider the case where the observer rotates around the acceleration direction. We expect that the OAM spectrum will be different, because when changing the sign of the azimuthal angle $\theta$, the metric is not invariant. Moreover, there is a cross term between the temporal and spatial elements, which we expect to induce some modifications in wave functions. We will further investigate OAM entanglement in this frame to see how rotation motion affects it.

Unless otherwise specified, geometrized units with $c=G=k_B=1$ are used. The metric signature is chosen as $(-,+,+,+)$. All greek indices run in $\{0, 1, 2, 3\}$. This paper is structured as follows: In Sec. II, the spacetime structure of the stationary rotating accelerated frame is studied. Section III reviews the quantization of the scalar field in cylindrical coordinates and in the Rindler coordinates. In Sec. IV, the Bogoliubov transformation between the Minkowski and Rindler operators is identified and generalized to a single-mode transformation. In Sec V, the OAM spectrum of Rindler particles is studied by expressing the Minkowski vacuum with the Rindler modes. To further understand how the observer will experience the Rindler particles, in Sec. VI, the Unruh-DeWitt detector and its detailed balance relation are studied. Then, the OAM entanglement in the rotating accelerated frame is explored in Sec. VII. Last, we generalize our result to all rotating accelerated trajectories in Sec. VIII. Our results are summarized and discussed in Sec. IX.

\section{Stationary rotating accelerated frame}
As stated in the Introduction, a stationary trajectory can be generated by the combination of generators of the Poincar\'e group. For the main part of the paper, we will focus on a special case where the rotation is parallel to the acceleration. In the end, we will show that all the results obtained in this case can be generalized to all stationary trajectories. Now, let us consider an observer who is moving along the $z$ direction with an acceleration of $a \hat {\mathbf z}$ while rotating around the same axis with an angular velocity of $\omega \hat {\mathbf z}$. The tangent vectors of the observer's world line, which can be chosen to be the proper time vector of the observer, form a Killing field \cite{korsbakken2004fulling},\begin{equation}
\partial_\tau \equiv \mathbf v=\mathbf {P_t}+\omega \mathbf {J_z}+a\mathbf{K_z},\label{eq.KillingField}
\end{equation}where $\partial_\tau$ is the proper time vector, and $\mathbf {P_t}=\partial_t$, $\mathbf {J_z}=(x\partial_y-y\partial_x)$, and $\mathbf{K_z}=(z\partial_t+t\partial_z)$ generate the time translation, the rotation around the $z$ axis, and the boost along the $z$ axis, respectively. When the observer moves along $\mathbf v$, the spacetime metric in the observer's rest frame will be constant. Thus, we say the observer is moving along a stationary trajectory. By defining $z=z'-1/a$, the Killing field becomes \begin{equation}
\partial_\tau=\omega \mathbf {J_{z'}}+a\mathbf{K_{z'}},
\end{equation}
where $\mathbf {P_t}$ is absorbed into $\mathbf{K_{z'}}$. We can further introduce the following coordinates in the $z'-t$ plane and $x-y$ plane, respectively \cite{korsbakken2004fulling}:
\begin{align}
z'=&\xi \cosh \chi,&t=&\xi\sinh\chi;\\
x=&r \cos \theta ,&y=&r\sin\theta.
\end{align}
The coordinates $(\chi,~\xi,~r,~\theta)$ are called the Rindler coordinates, and $\chi$ is the time coordinate. Now, the Killing field can be further simplified into \begin{equation}
\partial_\tau=\omega \partial_\theta+a\partial_\chi, \label{eq.stationaryTrajectory}
\end{equation}
by which the worldline of the observer can be calculated. Suppose at $\tau=0$, the observer is located at $t_{\rm{obs}}(\tau)=0,~x_{\rm{obs}}(\tau)=r_0,~y_{\rm{obs}}(\tau)=0,~z_{\rm{obs}}(\tau)=0$. Then, the worldline of the observer is given by\begin{align}
\chi_{\rm{obs}}(\tau)=&a\tau,&\xi_{\rm{obs}}(\tau)=&\frac 1 a, \label{eq.worldline1}\\
r_{\rm{obs}}(\tau)=&r_0,&\theta_{\rm{obs}}(\tau)=&\omega\tau,\label{eq.worldline2}
\end{align}
where $\chi_{\rm{obs}}(\tau)$ and $\xi_{\rm{obs}}(\tau)$ describe the accelerating motion along the $z'$ axis, while $r_{\rm{obs}}(\tau)$ and $\theta_{\rm{obs}}(\tau)$ describe the rotation around the $z'$ axis.

The Minkowski metric, in the Rindler coordinates, is given by \begin{equation}
ds^2=-\xi^2 d\chi^2+d\xi^2+dr^2+r^2 d\theta^2.
\end{equation}
At $\xi=0$, for an arbitrary nonzero vector $u^\mu$, its norm is $g_{\mu\nu}u^\mu u^\nu >0.$ That means that all nonzero vectors are spacelike, and therefore, no timelike or null curves can penetrate the $\xi=0$ plane. Thus, $\xi=0$ are the horizons that split the spacetime into a right region and a left region. These regions are spacelike to each other, so an observer on one side can never acquire the information on the other side.

Unlike the nonrotating case, there exists an extra special structure. This can be seen by the norm of the Killing field, which is given by $g_{\mu\nu}v^\mu v^\nu=\omega^2r^2-a^2\xi^2.$ Setting the norm to zero, we have $r=\pm a\xi /\omega$, beyond which $g_{\mu\nu}v^\mu v^\nu$ is positive and $\mathbf v$ becomes spacelike, so the observer moving along $\mathbf v$ cannot stay static in the three-dimensional space. Thus, $r_*=\pm a\xi /\omega$ is the static limit, as in the spacetime near a rotating black hole. The existence of the static limit complicates the quantization procedure because the "timelike" Killing vector $v^\mu$ is only timelike outside the static limit. It is not clear how to deal with the field modes inside the static limit, i.e., $r>r_*$. There have been several methods proposed to tackle this problem. The first method assumes the usual quantization procedure in the static spacetimes, i.e., choosing mode functions whose Killing time dependence will continue to hold inside the static limit \cite{dewitt1975quantum,unruh1974second}. This is justified by the fact that there is a Killing field that is timelike inside the static limit and whose affine parameter is the same as that for the usual Killing field. The second method regards the spacelike hypersurfaces of constant Killing time as fundamental, not the timelike Killing field. This procedure can be viewed as "untwisting" the Killing field to get a new vector field which is orthogonal to a spacelike hypersurface, and then quantizing fields with respect to observers moving along integral curves of the new vector field \cite{dray1992scalar,ashtekar1975quantum}. Another approach circumvents this problem by eliminating the area inside the static limit altogether by imposing "perfect conductor" (i.e., vanishing) boundary conditions \cite{manogue1987vacuum,davies1996detecting}. In this paper, we combine the last two methods. We will first quantize the scalar field on some hypersurface with a constant time coordinate, i.e. $t$ or $\chi$. Then, the boundary condition at $r=r_*$ is imposed so that the quantization procedure is confined outside the static limit.

\section{Quantization of the Klein-Gordon field}
First, let us consider the Klein-Gordon equation in the cylindrical coordinates, which is given by \begin{equation}
\Box \psi=\left (-\partial^2_t+\partial^2_z+\partial^2_r+\frac 1 {r^2}\partial^2_\theta+\frac 1 r \partial_r \right )\psi=0,
\end{equation}
where $\psi$ is the Klein-Gordon field. The solutions to the equation are \begin{equation}
g_{lEk_3}(t,z,r,\theta)=C_{lEk_3}e^{il\theta-iEt+ik_3 z}J_l(Pr), \label{eq.modeInMinkowski}
\end{equation}where $C_{lEk_3}$ is a normalization constant, $l$ is an integer, $|k_3|<|E|$, $P=\sqrt{E^2-k^2_3}$, and $J_l(Pr)$ signifies the Bessel functions of the first kind. For now, $E$ can be any real value. As usual, $l$, $E$, and $k_3$ represent the azimuthal index (also known as the topological charge), the energy (or equivalently the frequency), and the $z$ component of the wave vector, respectively. We will refer to Eq. \eqref{eq.modeInMinkowski} as the Minkowski modes.

We will use Eq. \eqref{eq.innerProduct} as the inner product throughout this paper. For the Minkowski modes, the indices in Eq. \eqref{eq.innerProduct}, $I$ and $J$, represent the collection of $lEk_3$. By decomposing the four-dimensional spacetime into time and three-dimensional space, the spacetime metric $g_{\mu\nu}$ becomes \cite{misner1973gravitation}\begin{equation}
g_{\mu\nu}=\begin{pmatrix}
-N^2+\beta_k \beta^k & \beta_j \\
\beta_i &\gamma_{ij}
\end{pmatrix},
\end{equation}
where $N$ is the lapse function, $\beta_i$ is the shift vector, $\beta^j=\gamma^{ij}\beta_i$, and the unit normal vector is given by $n^\mu=(1/N,-\beta^1/N,-\beta^2/N,-\beta^3/N)$. Since proper wave functions should be zero at infinity, the Stokes theorem ensures that the integration \eqref{eq.innerProduct} is independent of the hypersurface $\Sigma$. For most cases, it would be easier to calculate if the hypersurface $\Sigma$ is chosen to be the hyperplane $x^0=0$. Then, the inner product for the two Minkowski modes $g_{lEk_3}(t,z,r,\theta)$ and $g_{l'E'k'_3}(t,z,r,\theta)$ is given by \begin{align}
&~~\left < g_{lEk_3}(t,z,r,\theta),g_{l'E'k'_3}(t,z,r,\theta)\right >\nonumber \\&=\frac {8 \pi^2 E |C_{lEk_3}|^2}{P}\delta(l-l')\delta(k_3-k'_3)\delta(P-P'), \label{eq.innerProductForMinkowskiMode}
\end{align}
where the closure equation of the Bessel functions is used, i.e., \begin{equation}
\int_0^\infty  r J_l(Pr)J_l(P'r)dr=\frac 1 P \delta(P-P').
\end{equation}
Therefore, for positive-norm modes, we should choose $E>0$ on the whole Minkowski spacetime. Since $E$ and $P$ are both positive, one of them can be determined if we know the other one. Therefore, we may sometimes write $g_{lPk_3}(t,z,r,\theta)$ instead of $g_{lEk_3}(t,z,r,\theta)$, when it appears to be more convenient. From Eq. \eqref{eq.innerProductForMinkowskiMode}, we also know the normalization constant is given by \begin{equation}
C_{lEk_3}=\left (\frac P{8\pi^2 E} \right )^{1/2}=\frac {(E^2-k^2_3)^{1/4}}{(8\pi^2 E)^{1/2}}.
\end{equation}
After canonically quantizing the Klein-Gordon field, we may write the field operator as \begin{equation}
\psi(x^\mu)=\sum_{l,E,k_3} \left [a_{lEk_3}g_{lEk_3}(x^\mu)+a^\dagger_{lEk_3}g^*_{lEk_3}(x^\mu)\right ],\label{eq.psiMinkowski}
\end{equation} 
where $a_{lEk_3}$ and $a^\dagger_{lEk_3}$ are the usual annihilation and creation operators, respectively, and in this paper, we omit the hats on operators when no confusion arises. The summation symbol means that the indices run over all possible values. Since $E$ and $k_3$ vary continuously, the summation should be replaced by integration, but for simplicity, when no confusion is caused, we use the summation symbol instead. One may also verify the following commutation relations: \begin{align}
\big [a_{lEk_3},a^\dagger_{l'E'k'_3} \big ]&=\delta(l-l')\delta(E-E')\delta(k_3-k'_3),\\
\big [a_{lEk_3},a_{l'E'k'_3} \big ]&=\big [a^\dagger_{lEk_3},a^\dagger_{l'E'k'_3} \big ]=0,
\end{align}
and the number operator is given by $N_{lEk_3}=a^\dagger_{lEk_3}a_{lEk_3}$.

Next, we will try to quantize the Klein-Gordon field in the Rindler coordinates. From the Killing field, Eq. \eqref{eq.stationaryTrajectory}, we can write \begin{equation}
\hat H_{\rm{RF}}=-\omega \hat  J_{\mathbf {z'}}-a\hat  K_{\mathbf {z'}}, \label{eq.HRF}
\end{equation}
where $\hat H_{\rm{RF}}=i\partial_\tau$, $ \hat  J_{\mathbf {z'}}=-i(x\partial_y-y\partial_x)=-i\partial_\theta$, and $\hat  K_{\mathbf {z'}}=-i(z'\partial_t-t\partial_{z'})=-i\partial_\chi$ are the Hamiltonian operator, the $z'$ component of the angular momentum operator, and the boost operator along the $z'$ axis, respectively. Thus, the Hamiltonian $\hat H_{\rm{RF}}$ is defined in the rest frame of the observer, and it will be independent of time. This fact is another reason that we say the observer is moving along a stationary trajectory. Given a Hamiltonian, $H_{\rm{IF}}=i\partial_0$, defined in any inertial frame, one may derive that \begin{equation}
\left [H_{\rm{IF}},K_{\mathbf {z'}} \right ]=iP_{\mathbf {z'}},
\end{equation}
where $ P_{\mathbf {z'}}=-i\partial_{z'}$ is the linear momentum operator along the $z'$ axis. The nonzero commutator indicates that the commutator $\left [H_{\rm{IF}},H_{\rm{RF}} \right ]$ is also nonzero. Therefore, the particle states defined by $H_{\rm{IF}}$ and $H_{\rm{RF}}$ are different. In particular, the vacuum states defined by them are not equivalent. The vacuum state defined by $H_{\rm{IF}}$ is called the Minkowski vacuum state, while that defined by $H_{\rm{RF}}$ is the Fulling-Rindler vacuum. These two vacuum states are the only vacua that can appear along a stationary trajectory in Minkowski spacetime \cite{letaw1981quantized}.

It would be easier to find the eigenstates of $H_{\rm RF}$ in the Rindler coordinates, and then infer their energy by using Eq. \eqref{eq.HRF}. Let us denote the eigenfunctions of $H_{\rm{RF}}$ in Rindler coordinates as $h_{m\Omega}(\chi,\xi,r,\theta)$, where $m$ and $\Omega$ are related to the eigenvalues of $J_{\mathbf {z'}}$ and $K_{\mathbf {z'}}$, respectively, i.e.,\begin{align}
J_{\mathbf {z'}}h_{m\Omega}(\chi,\xi,r,\theta)=&m h_{m\Omega}(\chi,\xi,r,\theta),\\
K_{\mathbf {z'}}h_{m\Omega}(\chi,\xi,r,\theta)=& -\Omega h_{m\Omega}(\chi,\xi,r,\theta).
\end{align}
The reason that $\Omega$ is defined in this way is that later we require $\Omega>0$ for the positive-norm modes. Hence, we can write \begin{equation}
h_{m\Omega}(\chi,\xi,r,\theta)=H_{m\Omega}(\xi,r)e^{-i\Omega \chi+i m \theta},\label{eq.hmomega}
\end{equation}
where $H_{m\Omega}(\xi,r)$ is a function to be determined by the Klein-Gordon equation, which in the Rindler coordinates is given by \begin{equation}
\left (-\frac {\partial^2_\chi} {\xi^2}+\frac {\partial_\xi} \xi +\partial^2_\xi+\partial^2_r+\frac {\partial^2_\theta} {r^2} +\frac {\partial_r} r  \right )h_{m\Omega}(\chi,\xi,r,\theta)=0.
\end{equation} 
Substituting Eq. \eqref{eq.hmomega} into it gives \begin{equation}
\left (\partial^2_\xi+\frac {\partial_\xi} \xi +\partial^2_r+\frac {\partial_r} r+\frac{\Omega^2}{\xi^2}-\frac {m^2}{r^2}   \right )H_{m\Omega}(\xi,r)=0.
\end{equation}
The solutions are proportional to $J_m(Qr)K_{i\Omega}(|Q\xi|) $, where $K_{\alpha}(x)$ are the modified Bessel functions of the second kind, and $Q$ is a new parameter. Therefore, the mode functions, named the Rindler modes, can be written as\begin{equation}
h^{(\sigma)}_{m\Omega Q}(\chi,\xi,r,\theta)=D^{(\sigma)}_{m\Omega Q} e^{-i\Omega \chi+i m \theta}J_m(Qr)K_{i\Omega}(\sigma Q\xi),\label{eq.RindlerMode}
\end{equation}
where $D_{m\Omega Q}$ is the normalization constant, and $\sigma=+$ means the mode is defined in the right region with $\xi>0$, while $\sigma=-$ means the opposite. The azimuthal index $m$ can take any integer. Unlike the Minkowski mode functions \eqref{eq.modeInMinkowski} in cylindrical coordinates, $Q$ is not related to other indices, and it is only required that $Q>0$. For now, $\Omega$ can be any real value. Since $\Omega$ appears before the Rindler time coordinate $\chi$, it will be referred to as the Rindler energy. However, it is not the energy perceived by the observer. From the perspective of the observer, the energy of a particle $\mathcal E$ should satisfy \begin{equation}
H_{\rm{RF}} h(\chi,\xi,r,\theta)=\mathcal E h(\chi,\xi,r,\theta).\label{eq.definitionOfE}
\end{equation}
Therefore, we see that the energy of a particle is related to the Rindler energy by \begin{equation}
\mathcal E=a\Omega-m\omega. \label{eq.energy}
\end{equation}
The inner product of the Rindler mode functions is given by\begin{align}
&\left <h^{(\sigma)}_{m\Omega Q}(\chi,\xi,r,\theta) ,h^{(\sigma')}_{m'\Omega' Q'}(\chi,\xi,r,\theta)\right >\nonumber \\=&\frac {2 \pi^3\sigma \left (D^{(\sigma)}_{m\Omega Q} \right )^2}{Q\sinh (\Omega \pi)}\delta (m-m')\delta (\Omega-\Omega')\delta (Q-Q')\delta (\sigma-\sigma').
\end{align}
Therefore, the normalization constant may be written as \begin{equation}
D^{(\sigma)}_{m\Omega Q}=\sqrt{\frac{Q \sinh(\sigma \Omega \pi)}{2 \pi^3}}.
\end{equation}
Also, in the right region, where $\xi>0$ and $\sigma=+$, we should choose $\Omega>0$ for positive-norm modes, while $\Omega<0$ in the left region. This is different from the Minkowski modes \eqref{eq.modeInMinkowski} in the cylindrical coordinates, where $E>0$ in both regions. There is another interesting fact in this case. In our earlier work \cite{wu2023orbital}, where the OAM particles in the linear accelerated frame are studied, the particles in the right region all have positive-defined energy. However, when the observer starts to rotate, things change. Now, the particles in the right region can also have negative energy whenever $a\Omega<m\omega$, which can be seen from Eq. \eqref{eq.energy}. We will explore this phenomenon further later. Similarly, we can write the Klein-Gordon field by the Rindler modes as \begin{equation}
\psi(x^\mu)=\sum_{m,\Omega,Q,\sigma} \left [b^{(\sigma)}_{m\Omega Q}h^{(\sigma)}_{m\Omega Q}(x^\mu)+b^{(\sigma)\dagger}_{m\Omega Q}h^{(\sigma)*}_{m\Omega Q}(x^\mu)\right ],\label{eq.psiRindler}
\end{equation}
where $b^{(\sigma)}_{m\Omega Q}$ and $b^{(\sigma)\dagger}_{m\Omega Q}$ are the usual annihilation and creation operators for the Rindler modes, respectively. They also obey the usual commutation relations,
\begin{align}
\big [b^{(\sigma)}_{m\Omega Q},b^{(\sigma')\dagger}_{m'\Omega' Q'} \big ]&=\delta (m-m')\delta (\Omega-\Omega')\delta (Q-Q')\delta (\sigma-\sigma'),\\
\big [b^{(\sigma)}_{m\Omega Q},b^{(\sigma')}_{m'\Omega' Q'} \big ]&=\big [b^{(\sigma)\dagger}_{m\Omega Q},b^{(\sigma')\dagger}_{m'\Omega' Q'}\big ]=0. 
\end{align}

The Minkowski modes \eqref{eq.modeInMinkowski} and the Rindler modes \eqref{eq.RindlerMode} are well defined on the spacelike hypersurfaces with constant $t$ and $\chi$, respectively. Hence, they are well defined with respect to nonrotating observers. However, this will not hold when the observers start to rotate, because of the existence of the static limit. As stated in the last section, after quantizing the scalar field on some spacelike hypersurface, we now need to impose the vanishing boundary condition to confine the scalar field outside the static limit, and the integrals with respect to $r$ are restricted from $0$ to $r_*$. For the modes in Eq. \eqref{eq.modeInMinkowski}, this can be done by using $J_l(\tilde u_{l,P} r)$ to replace $J_l(Pr)$, where $\tilde u_{l,P}=u_{l,P}/r_*$, and $u_{l,P}$ is the $P$th zero of the Bessel function $J_l(x)$. This replacement also makes the continuous label $P$ become discrete. Now, it can only take positive whole numbers. Accordingly, the normalization constant is replaced by $C_{lEk_3}=[2 \pi \sqrt{E}r_*\left | J_{l+1}(u_{l,P}) \right | ]^{-1}$ and $E^2=k^2_3+\tilde u_{l,P}^2$. Similarly, for the modes in Eq. \eqref{eq.RindlerMode}, the Bessel functions $J_m(Qr)K_{i\Omega}(\sigma Q\xi) $ are replaced by $J_m(\tilde u_{m,Q} r)K_{i\Omega}(\sigma \tilde u_{m,Q} \xi) $ and $Q$ can only take positive whole numbers, as well. The normalization constant is $D^{(\sigma)}_{m\Omega Q}=\pi^{-3/2}[\sinh (\sigma \Omega \pi)]^{1/2}/[r_* J_{m+1}(u_{m,Q})]$.

\section{Single-mode Bogoliubov transformation}
From the expansions for the field operator, Eqs. \eqref{eq.psiMinkowski} and \eqref{eq.psiRindler}, one can see that the Bogoliubov transformation, which connects the Minkowski and Rindler operators, is given by \begin{align}
b^{(\sigma)}_{m\Omega Q} =&\sum_{l,P,k_3} \left ( \alpha ^{(\sigma)*}(m,\Omega,Q;l,P,k_3) a_{l,P,k_3}\right .\nonumber \\&\left .-\beta ^{(\sigma)*}(m,\Omega,Q;l,P,k_3) a^\dagger_{l,P,k_3} \right ),
\end{align}
where the Bogoliubov coefficients are given by \begin{align}
&\alpha ^{(\sigma)}(m,\Omega,Q;l,P,k_3)=\left <g_{lPk_3}(x^\mu), h^{(\sigma)}_{m\Omega Q}(x^\mu)\right >\nonumber \\ =&D^{(\sigma)}_{m\Omega Q}C_{lPk_3}\frac {\pi^2 e^{\sigma \Omega \pi/2}\left [r_*J_{m+1}(u_{m,Q}) \right ]^2 e^{ik_3/a}} {\sinh \left (\sigma \Omega \pi\right )} \nonumber \\ &\times\left ( \frac {E-k_3}{E+k_3}\right )^{i\Omega/2} \delta(m-l)\delta(Q-P) , \\
&\beta ^{(\sigma)}(m,\Omega,Q;l,P,k_3)=-\left <g^*_{lPk_3}(x^\mu), h^{(\sigma)}_{m\Omega Q}(x^\mu)\right >\nonumber \\ =&D^{(\sigma)}_{m\Omega Q}C_{lPk_3}(-1)^m\frac {\pi^2 e^{-\sigma \Omega \pi/2}\left [r_*J_{m+1}(u_{m,Q}) \right ]^2 e^{-ik_3/a} } {\sinh\left (-\sigma \Omega \pi\right )}\nonumber \\ &\times \left ( \frac {E-k_3}{E+k_3}\right )^{i\Omega/2}\delta(m+l)\delta(Q-P) .
\end{align}
Since we mainly consider an observer moving in the right region, we will restrict $\Omega>0$ from now on, and replace all $\Omega$'s with $\sigma \Omega$. Now, the Bogoliubov transformation for $b^{(\sigma)}_{m\Omega Q}$ is given by\begin{align}
b^{(\sigma)}_{m\Omega Q} =&\int dk_3 \bigg [   \frac {e^{-ik_3/a}  e^{ \Omega \pi/2}}{2\sqrt{\pi E \sinh(\Omega \pi)}}\left ( \frac {E-k_3}{E+k_3}\right )^{-i\sigma \Omega/2} a_{mQk_3}\nonumber \\&+ \frac {(-1)^m e^{ik_3/a}  e^{- \Omega \pi/2}}{2\sqrt{\pi E \sinh(\Omega \pi)}}\left ( \frac {E-k_3}{E+k_3}\right )^{-i\sigma \Omega/2} a^\dagger_{-mQk_3} \bigg ].\label{eq.Btransform}
\end{align}
The Bogoliubov transformation for $b^{(\sigma)\dagger}_{m\Omega Q}$, $a_{l,P,k_3}$, and $a^\dagger_{l,P,k_3}$ can also be derived. The transformation can be used to study the Minkowski vacuum state and one-particle state in the Rindler coordinates, but the derivation will be messy. Instead, since we mainly care about the degrees of freedom of OAM, we shall find a single-mode Bogoliubov transformation as follows \cite{takagi1986vacuum}.

First, Eq. \eqref{eq.Btransform} inspires us to define\begin{equation}
P^{(\sigma)}_{\Omega}(k_3)=\frac 1 {\sqrt{2\pi E}}e^{-ik_3/a}\left ( \frac {E-k_3}{E+k_3} \right )^{-i\sigma\Omega/2}.
\end{equation}
This function is orthonormal and complete, i.e.,\begin{align}
&\int_{-\infty}^\infty dk_3 P^{(\sigma)}_{\Omega}(k_3) P^{(\sigma)'}_{\Omega'}(k_3)=\delta(\sigma-\sigma')\delta(\Omega-\Omega'),\\
&\sum_\sigma \int_0^\infty d\Omega P^{(\sigma)}_{\Omega}(k_3)P^{(\sigma)*}_{\Omega}(k'_3)=\delta(k_3-k'_3).
\end{align}
Then, we can define a new set of annihilation and creation operators as\begin{align}
a^{(\sigma)}_{m\Omega Q}=&\int_{-\infty}^\infty dk_3 P^{(\sigma)}_{\Omega}(k_3) a_{mQk_3},\label{eq.aNew}\\
a^{(\sigma)\dagger}_{m\Omega Q}=&\int_{-\infty}^\infty dk_3 P^{(\sigma)*}_{\Omega}(k_3) a^\dagger_{mQk_3}.\label{eq.adaggerNew}
\end{align}
Here, $a_{mQk_3}$ and $a^\dagger_{mQk_3}$ are operators corresponding to the Minkowski modes, Eq. \eqref{eq.modeInMinkowski}, while $a^{(\sigma)}_{m\Omega Q}$ and $a^{(\sigma)\dagger}_{m\Omega Q}$ are operators of a new set of Minkowski modes, denoted by $f^{(\sigma)}_{m\Omega Q}(x^\mu)$. Since $P^{(\sigma)}_{\Omega}(k_3)$ is orthonormal and complete, $f^{(\sigma)}_{m\Omega Q}$ are linear combinations of $g_{lPk_3}(x^\mu)$ and they share the same Minkowski vacuum. For the same reason, the new annihilation and creation operators satisfy the following commutation relation: \begin{align}
\big [a^{(\sigma)}_{m\Omega Q},a^{(\sigma')\dagger}_{m'\Omega' Q'} \big ]&=\delta (m-m')\delta (\Omega-\Omega')\delta (Q-Q')\delta (\sigma-\sigma'),\\
\big [a^{(\sigma)}_{m\Omega Q},a^{(\sigma')}_{m'\Omega' Q'} \big ]&=\big [a^{(\sigma)\dagger}_{m\Omega Q},a^{(\sigma')\dagger}_{m'\Omega' Q'}\big ]=0.
\end{align}
Later, when studying the one-particle state, we will find that when $a^{(+)\dagger}_{m\Omega Q}$ acts on the vacuum state, it will create a particle moving in the right region, while $a^{(-)\dagger}_{m\Omega Q}$ does the opposite. Now, the single-mode Bogoliubov transformation can be written as\begin{equation}
b^{(\sigma)}_{m\Omega Q}=\alpha^*(\Omega)a^{(\sigma)}_{m\Omega Q}-\beta^*(\Omega,-m)a^{(-\sigma)\dagger}_{-m\Omega Q},\label{eq.singleModeTransformation}
\end{equation}
where $\alpha(\Omega)=e^{\Omega \pi/2}/\sqrt{2\sinh(\Omega \pi )}$ and $\beta(\Omega,m)=(-1)^{m+1}e^{-\Omega \pi/2}/\sqrt{2\sinh(\Omega \pi )}=(-1)^{m+1}e^{-\Omega \pi}\alpha(\Omega)$. Note that there is no summation implied over repeated labels.

\section{OAM spectrum of Rindler particles in Minkowski vacuum}
Since the Bogoliubov coefficient $\beta$ is not zero, the Minkowski vacuum and the Fulling-Rindler vacuum will not be equivalent. According to the Unruh effect \cite{unruh1976notes,parikh2000hawking,visser2003essential,russo1992end,page2005hawking,robinson2005relationship,carroll2019spacetime,martin2010population,martin2010unveiling,alsing2003teleportation,fuentes2005alice,alsing2006entanglement,adesso2007continuous,ling2007quantum,takagi1986vacuum,wu2023orbital}, there exist Rindler particles in the Minkowski vacuum. Now, we will explore the OAM spectrum of these particles when the observer starts to rotate around the $z'$ axis. Let us denote the Minkowski vacuum state and the Fulling-Rindler vacuum state by $\left |0\right >_{\rm{M}}$, and $\left |0\right >_{\rm{R}}\left |0\right >_{\rm{L}}$, respectively, where $\rm{R}$ means the right region and $\rm{L}$ the other one, so the operators with $\sigma=+$ only act on $\left |0\right >_{\rm{R}}$, and those with $\sigma=-$ on $\left |0\right >_{\rm{L}}$. Suppose the Minkowski vacuum can be written as $\left |0\right >_{\rm{M}}=F\left (b^{(\sigma)\dagger}_{m\Omega Q} \right )\left |0\right >_{\rm{R}}\left |0\right >_{\rm{L}}$, where $F\left (b^{(\sigma)\dagger}_{m\Omega Q} \right )$ is a function of all creation operators for the Rindler modes. One may derive that \begin{equation}
F\left (b^{(\sigma)\dagger}_{m\Omega Q} \right )\propto \exp \left ( \sum_{m\Omega Q} (-1)^me^{-\Omega \pi} b^{(+)\dagger}_{m\Omega Q}   b^{(-)\dagger}_{-m\Omega Q}   \right ),
\end{equation}
and the Minkowski vacuum state can be written as \begin{align}
\left |0\right >_{\rm{M}}\propto &\prod_{m\Omega Q}\sum_{n_{m\Omega Q}=0}^{\infty}(-1)^{m n_{m\Omega Q}}e^{-\Omega \pi n_{m\Omega Q}}\left |n_{m\Omega Q}\right >_{\rm{R}}\nonumber \\&\times\left |n_{m\Omega Q}\right >_{\rm{L}},
\end{align}
where $\prod$ represents direct products, and $n_{m\Omega Q}$ is the number of particles in mode $h^{(+)}_{m\Omega Q}$. For consistency, when we write $\left |n_{m\Omega Q}\right >_{\rm{L}}$, we actually mean that there are $n_{m\Omega Q}$ particles with an OAM of $(-m)$ in the left region. One may notice that the above expression is not normalizable because the coefficients are independent of $Q$, which can be any positive whole number. This reflects the fact that the Minkowski vacuum and the Fulling-Rindler vacuum are not unitarily equivalent \cite{takagi1986vacuum}, i.e., one cannot be transformed from the other one by a unitary operator. The expression is only valid for each mode, and the most important information it carries is the relative frequency that each mode represents. Since we mainly care about the number of particles with OAM $m$, we may focus our attention only on the label $m$ and suppose the observer (detector) can only be excited by particles with energy $\mathcal E$. Then, we can write\begin{align}
\left |0\right >_{\rm{M}}=&C \sum_{\{n_m=0\}}^{\infty}(-1)^{\sum_l l n_{l}}e^{- \pi \sum_l (\mathcal E+l\omega) n_{l}/a}\left |\{ n_l \}\right >_{\rm{R}}\nonumber \\&\times\left |\{ n_l \}\right >_{\rm{L}},\label{eq.MinkowskiVacuum}
\end{align}
where $C=\sqrt{\prod_l (1-e^{-2\pi (\mathcal E+l\omega)/a})}$, and $\{ n_l \}$ is a set of particle numbers with a different OAM $l$.

What OAM modes are allowed? When the observer is linearly accelerated, all OAM modes are permitted in the Bogoliubov transformation, which is another reason that causes the Rindler particle number in the Minkowski vacuum state to diverge. However, when the observer starts to rotate, according to Eq. \eqref{eq.energy} and the restriction that the Rindler energy must be positive, we can see that if the angular velocity of the observer $\omega$ is positive, then only modes with $l>-\mathcal E/\omega$ are allowed, and when $\omega<0$, those with $l<-\mathcal E/\omega$ are allowed. We define the critical OAM $l_c$ to be the one that splits the allowed and the forbidden modes. Hence, we may write\begin{equation}
l_c=\begin{cases} 
\lceil -\frac {\mathcal E} \omega \rceil & {\rm{if~}}\omega>0{\rm{~and~}}\mathcal{E}\bmod\omega\ne 0,\\ 
-\frac {\mathcal E} \omega +1 & {\rm{if~}}\omega>0{\rm{~and~}}\mathcal{E}\bmod\omega= 0,\\ 
\lfloor -\frac {\mathcal E} \omega \rfloor & {\rm{if~}}\omega<0{\rm{~and~}}\mathcal{E}\bmod\omega\ne 0,\\ 
-\frac {\mathcal E} \omega -1& {\rm{if~}}\omega<0{\rm{~and~}}\mathcal{E}\bmod\omega= 0, \end{cases}
\end{equation}
where $\lceil\cdot\rceil$ and $\lfloor\cdot\rfloor$ are the ceiling and floor functions, respectively. Then, the normalization constant can be further given by $C=\sqrt{\left (e^{-2\pi(\mathcal E+l_c \omega)/a};e^{-2\pi |\omega|/a} \right )_{\infty}}$, where $\left (a;q\right )_{n} =\prod_{k=0}^{n-1}(1-aq^k)$ is the $q$-Pochhammer symbol.

We also can calculate the probability of finding $n_l$ particles with OAM $l$, where $l$ runs over all permitted values, which is given by\begin{equation}
P(\{n_l\})=C^2 \exp \left (-2\pi\frac {\sum_l (\mathcal E+l\omega)n_l} a \right ), \label{eq.pnl}
\end{equation}
where $l$ runs over all permitted OAM values. And the ensemble average of the total OAM for these Rindler particles is given by \begin{equation}
\left <L_z \right >=\sum_{\{n_l \}} \left [\left (\sum_{m}m n_m \right )P(\{n_l\})\right ]=\sum_{l}\frac {l}{e^{2\pi (\mathcal E+l\omega)/a}-1},\label{eq.lz}
\end{equation}
where, as usual, $l$ and $m$ can take all permitted values.

However, there is a paradox. From the perspective of the Rindler observer who is moving along a rotating accelerated trajectory, the scalar field may not be in its ground state. Instead, the field may contain some particles and the detector is excited by these particles to pick up some energy and OAM. Meanwhile, the field decays to a lower energy state and loses some OAM. However, this would be absurd from the eye of an inertial observer, who, in turn, would see that the scalar field is already in its ground state, i.e., the vacuum state, and there are no particles. No more energy or OAM can be extracted from the field. It would seem that the conservation of energy and OAM is broken, because the Rindler observer gains some energy and OAM, while the field cannot decay to anything else. This paradox may be solved by noticing that the Rindler observer is accelerating and rotating, which means there exists some external source to maintain its acceleration and rotation \cite{birrell1984quantum,carroll2019spacetime}. Hence, the Rindler observer takes energy and OAM from the external source, instead of the field itself.

\begin{figure} [tbhp]
\centering
\subfloat[][$\mathcal E=1$, $l=3$.]{%
  \includegraphics[width=0.9\linewidth]{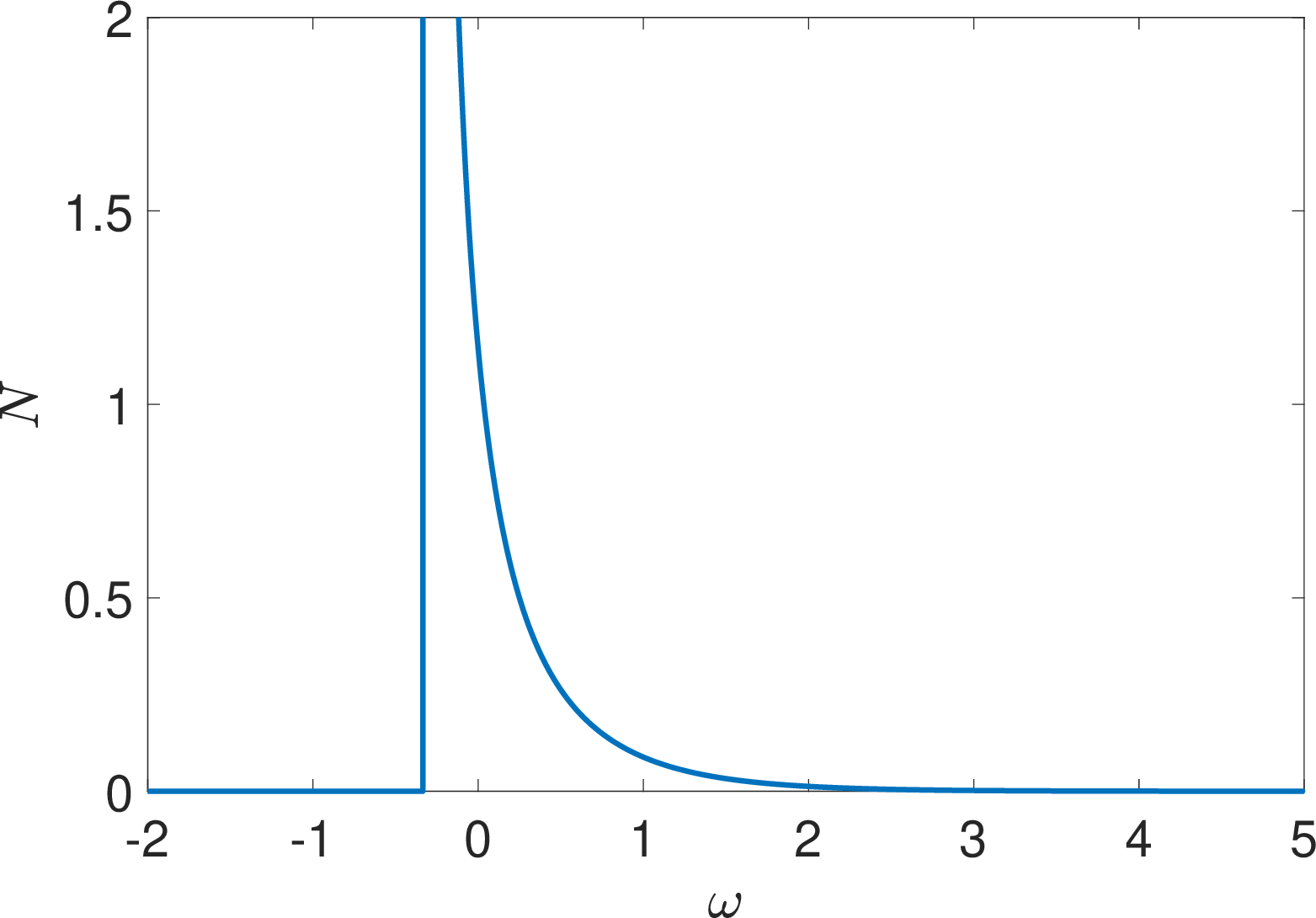}%
  \label{fig.np}
}\hfill
\subfloat[][$\mathcal E=-1$, $l=-3$.]{%
  \includegraphics[width=0.9\linewidth]{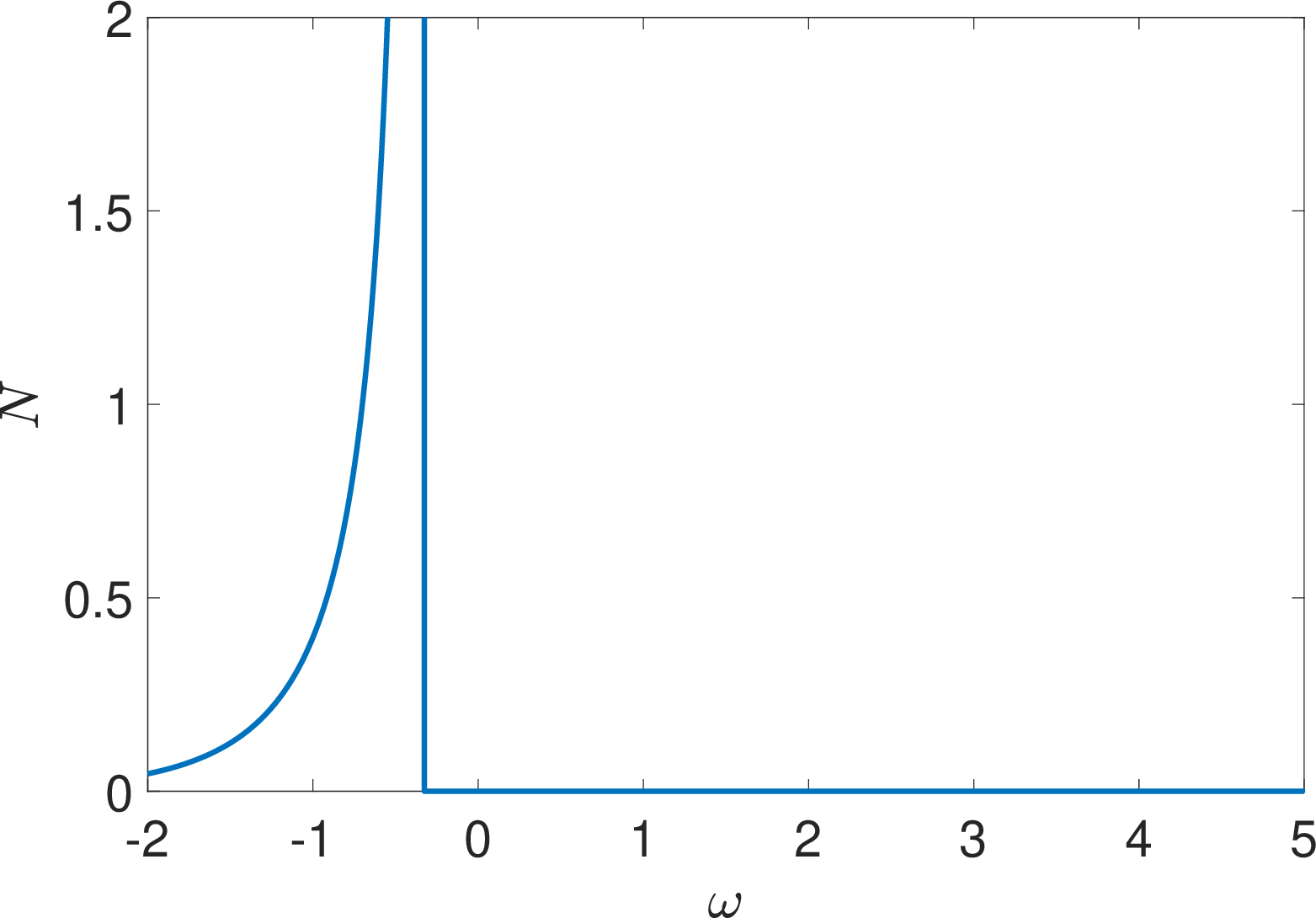}%
  \label{fig.nm}
}\hfill
\caption{The curves for expected particle number. The energy $\mathcal E$, angular velocity $\omega$, and the acceleration $a$ are in the units of $\rm J$, ${\rm J}/\hbar $, and $c{\rm J}/\hbar$, respectively. The acceleration $a$ is set to 10. The energy $\mathcal E$ and the OAM $l$ are, respectively, set to (a) $1$ and $3$, and (b) $-1$ and $-3$.}
\label{fig.n}
\end{figure}

The OAM spectrum is most easily seen by the expected number spectrum of Rindler particles, which is given by \begin{equation}
N_{m\mathcal E }=\prescript{}{\rm M}{\left <0\right |}b^{(\sigma)\dagger}_{m\mathcal E }b^{(\sigma)}_{m\mathcal E }\left |0 \right >_{\rm M}=\frac {e^{-(\mathcal E+m\omega)\pi/a}}{2\sinh((\mathcal E+m\omega)\pi/a)}. \label{eq.exNumParticle}
\end{equation}
As before, we omit the $Q$ label since it does not affect the spectrum. For later use, we first draw the curve for the expected particle number with fixed energy $\mathcal E=1$ and OAM $m=3$ by varying the angular velocity $\omega$ in Fig. \ref{fig.np}. The units of $\mathcal E$, $\omega$, and $a$ are $J$, $J/\hbar $, and $cJ/\hbar$, respectively. From the figure, we find that as $\omega$ increases from zero, the expected particle number decays exponentially. On the other hand, when $\omega \rightarrow -\mathcal E/m$ from the right, the expected particle number will approach infinity. As we will find out later, this infinite number of particles will cause entanglement to degrade to zero. If $\omega$ continues to decrease, when $\omega\le -\mathcal E/m$, no particle is allowed to appear, because otherwise the requirement that the Rindler energy must be positive will be violated. We also calculate the expected particle number for opposite energy and OAM, as in Fig. \ref{fig.nm}. This curve is conjugate to the above one. One can see that the particles with $\mathcal E=-1$ and $m=-3$ live in the region where those with $\mathcal E=1$ and $m=3$ are not allowed. As $\omega$ decreases from $-\mathcal E/m$, the expected particle number drops, as well.

\begin{figure} [tbhp]
\centering
\subfloat[][$l=0$.]{%
  \includegraphics[width=0.8\linewidth]{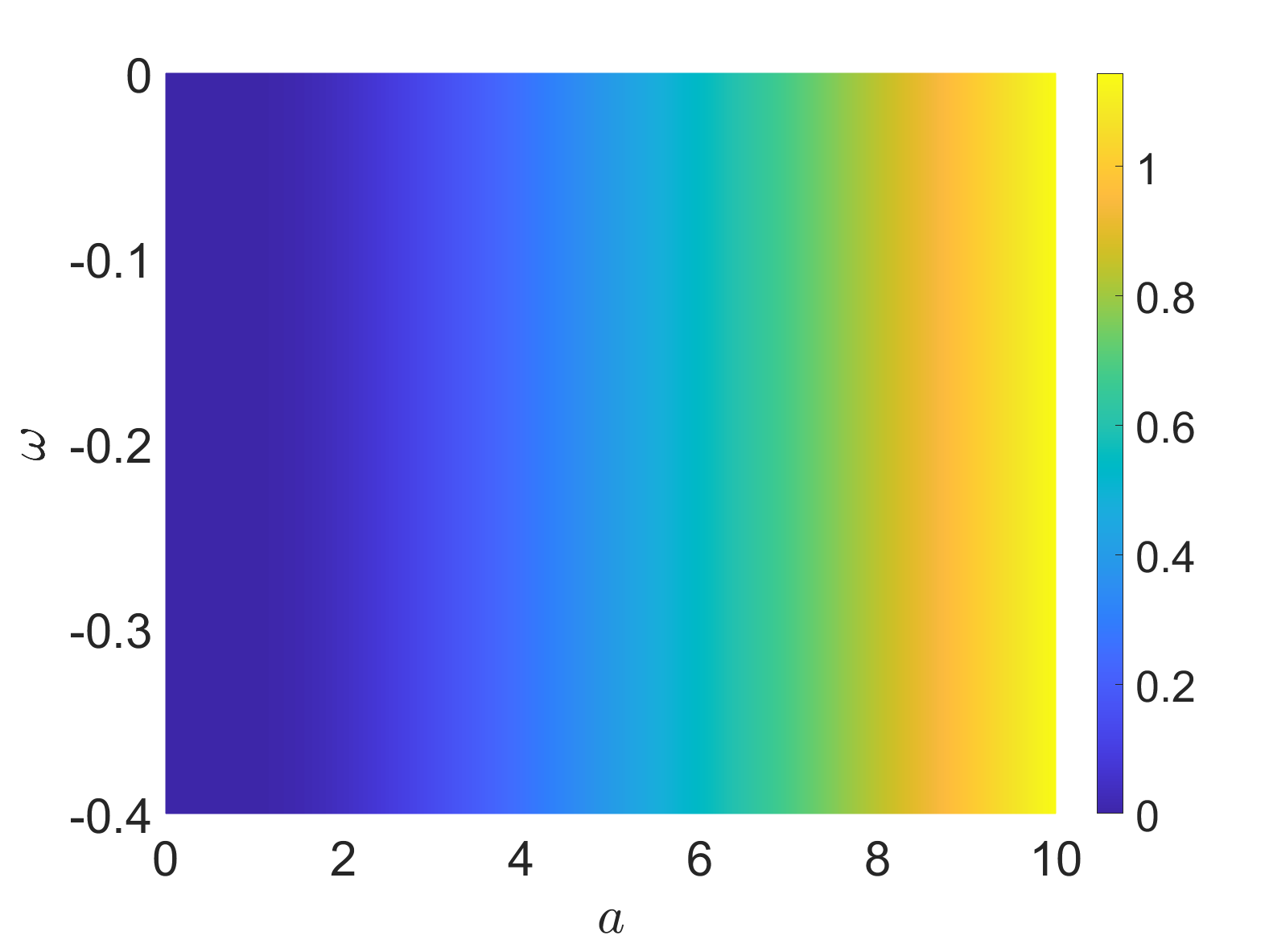}%
  \label{fig.aw0}
}\hfill
\subfloat[][$l=3$.]{%
  \includegraphics[width=0.8\linewidth]{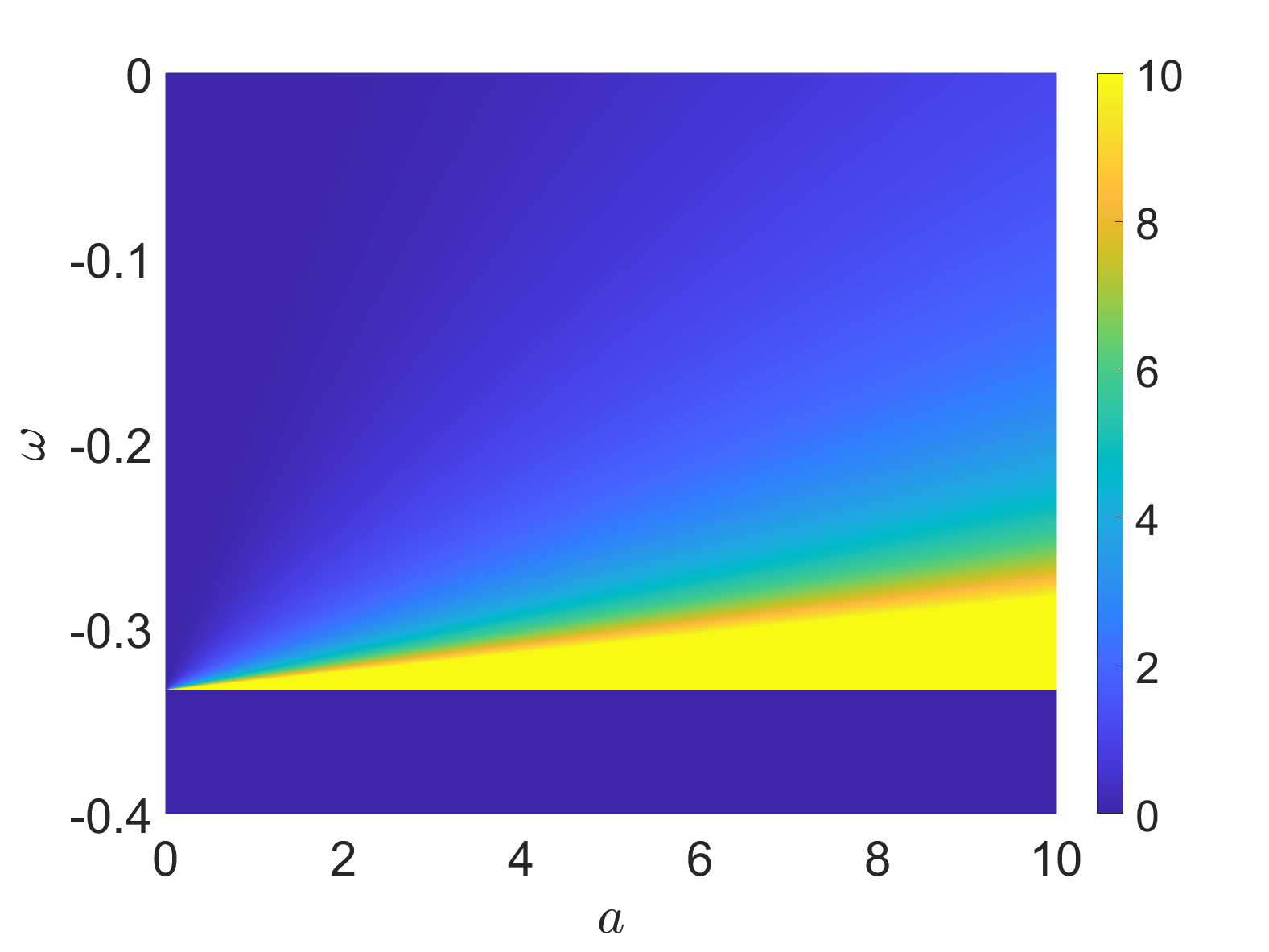}%
  \label{fig.aw3}
}\hfill
\subfloat[][$l=10$.]{%
  \includegraphics[width=0.8\linewidth]{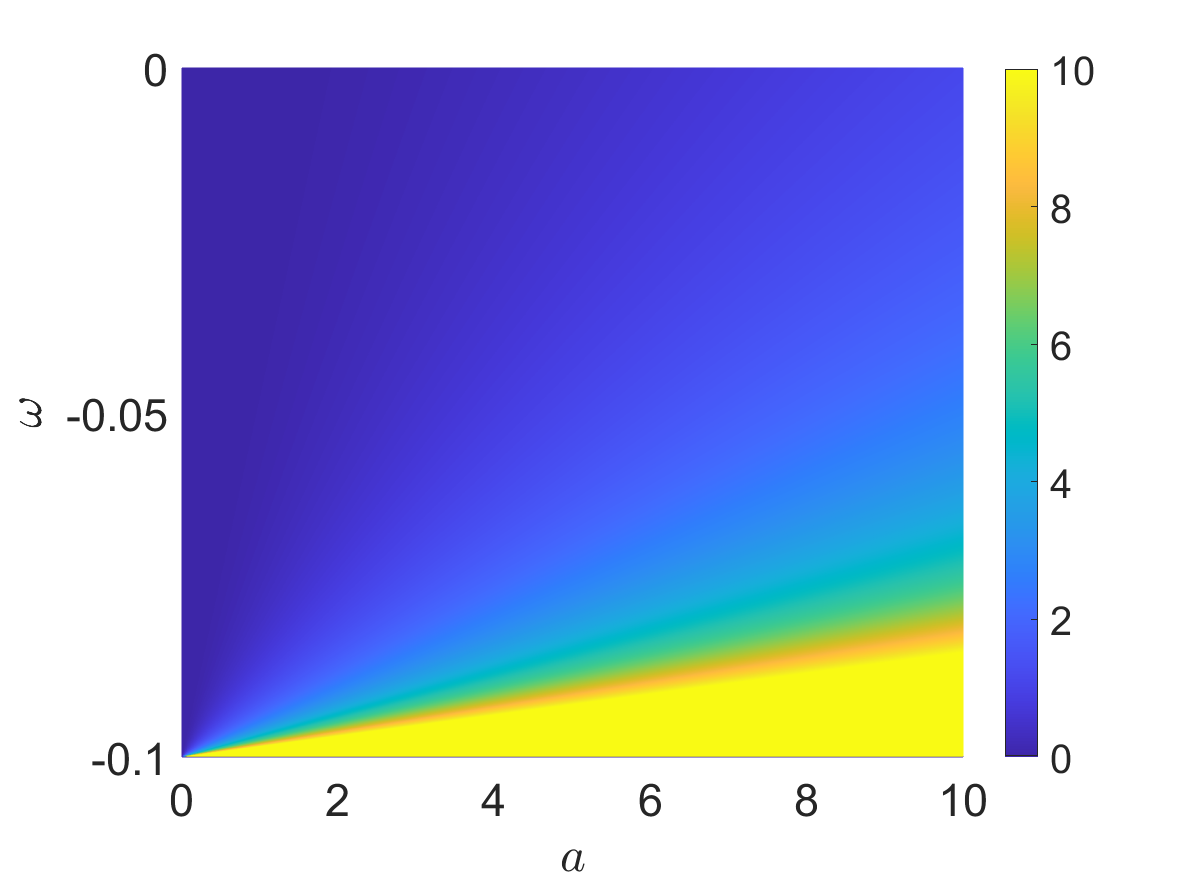}%
  \label{fig.aw10}
}\hfill
\caption{The expected particle numbers for (a) $l=0$, (b) $l=3$, and (c) $l=10$ modes. For better illustration, we truncate the particle number at 10 for $l=3$ and $l=10$ modes. The energy $\mathcal E$ is set to 1.}
\label{fig.aw}
\end{figure}

Currently, the detection of the Unruh effect is impossible, since it will require an enormously great acceleration. Many efforts have been focused on reducing the difficulties. For example, the experimentally verified Sokolov–Ternov effect may be related to the circular Unruh effect \cite{akhmedov2008physical}; the high-energy channeling radiation experiment could be the first observation of acceleration-induced thermality \cite{lynch2021experimental}; by using the geometric phase, it is shown that the acceleration needed can be as low as $10^{17}~{\rm m/s^2}$; a localized laser coupled to a Bose-Einstein condensate may be used to observe an analog of the circular Unruh effect \cite{gooding2020interferometric}. Meanwhile, Eq. \eqref{eq.exNumParticle} indicates that the rotation may facilitate the detection of Rindler particles when the OAM is nonzero. To find out how rotation affects the detection results, we draw the expected particle numbers with different acceleration $a$ and angular velocity $\omega$ for $l=0$, $l=3$, and $l=10$, respectively, in Fig. \ref{fig.aw}. In Fig. \ref{fig.aw0}, we can see that the rotation causes no changes in the expected particle numbers. This is because the energy shift term $m\omega$, in Eq. \eqref{eq.exNumParticle}, vanishes. In this case, the rotation will not reduce the difficulties of detecting Rindler particles. However, when the OAM is nonzero, as in Figs. \ref{fig.aw3} and \ref{fig.aw10}, the particle numbers are altered by the rotation. Note that we have truncated the particle numbers at 10 for better illustration. When the acceleration and the angular velocity have opposite directions, the expected particle numbers will be raised, before the angular velocity reaches a critical value $\omega_c={\mathcal E}/l$, where all particles will suddenly disappear. By comparing Fig. \ref{fig.aw3} with Fig. \ref{fig.aw10}, one can find that if we increase the OAM, then the same expected particle number can be achieved by using a smaller angular velocity. This is because by using a higher OAM $l$, the critical angular velocity $\omega_c$ moves towards zero, and the energy shift term $l\omega$ grows, as well. Therefore, we now have three degrees of freedom, i.e., the acceleration, the angular velocity, and the OAM value, all of which can enhance the detection as they increase (the angular velocity increases in the opposite direction).

\begin{figure} [tbhp]
\centering
\subfloat[][$\omega=0$.]{%
  \includegraphics[width=0.48\linewidth]{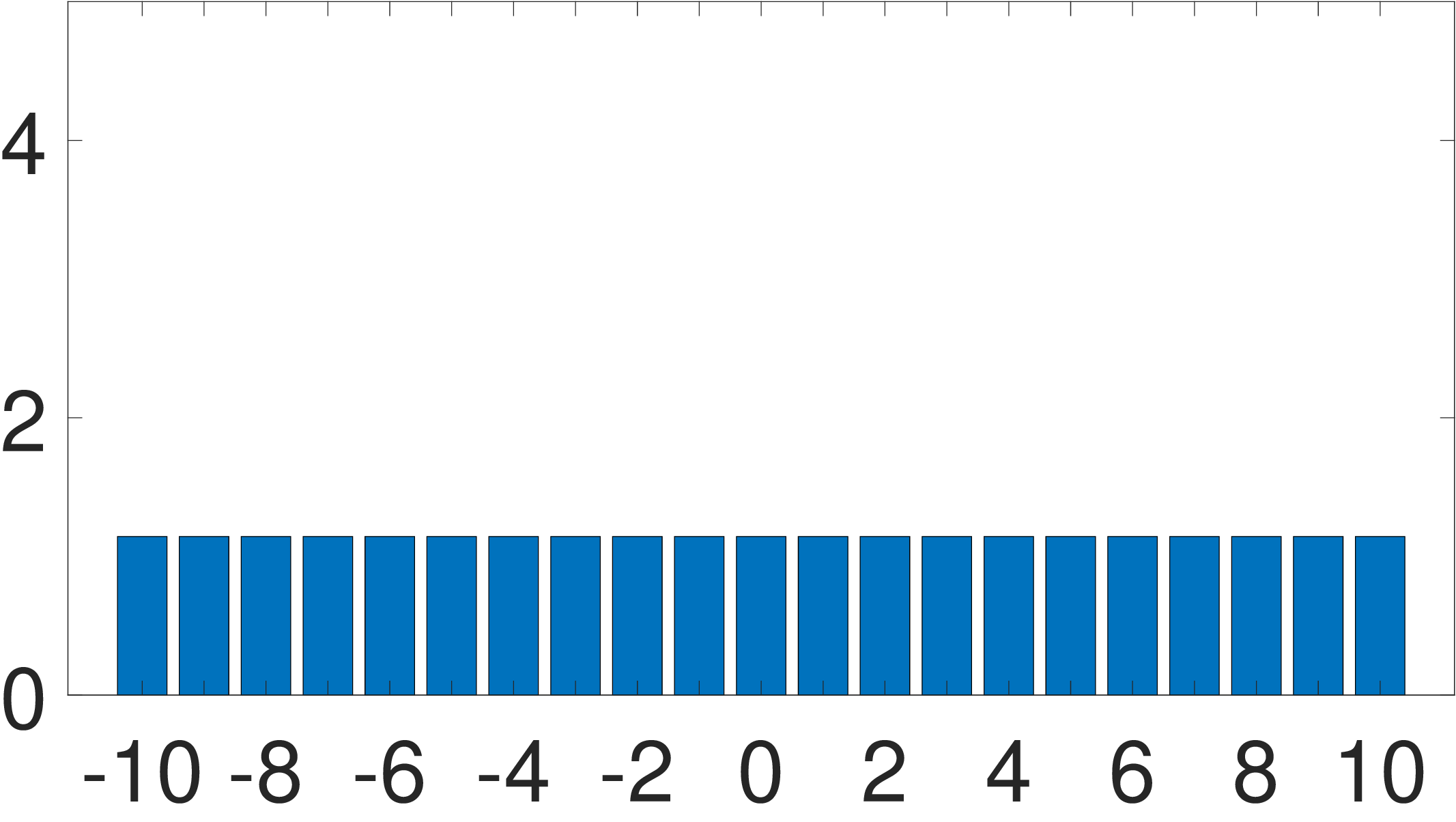}%
  \label{fig.p1p1}
}\hfill
\subfloat[][$\omega=0.1$.]{%
  \includegraphics[width=0.48\linewidth]{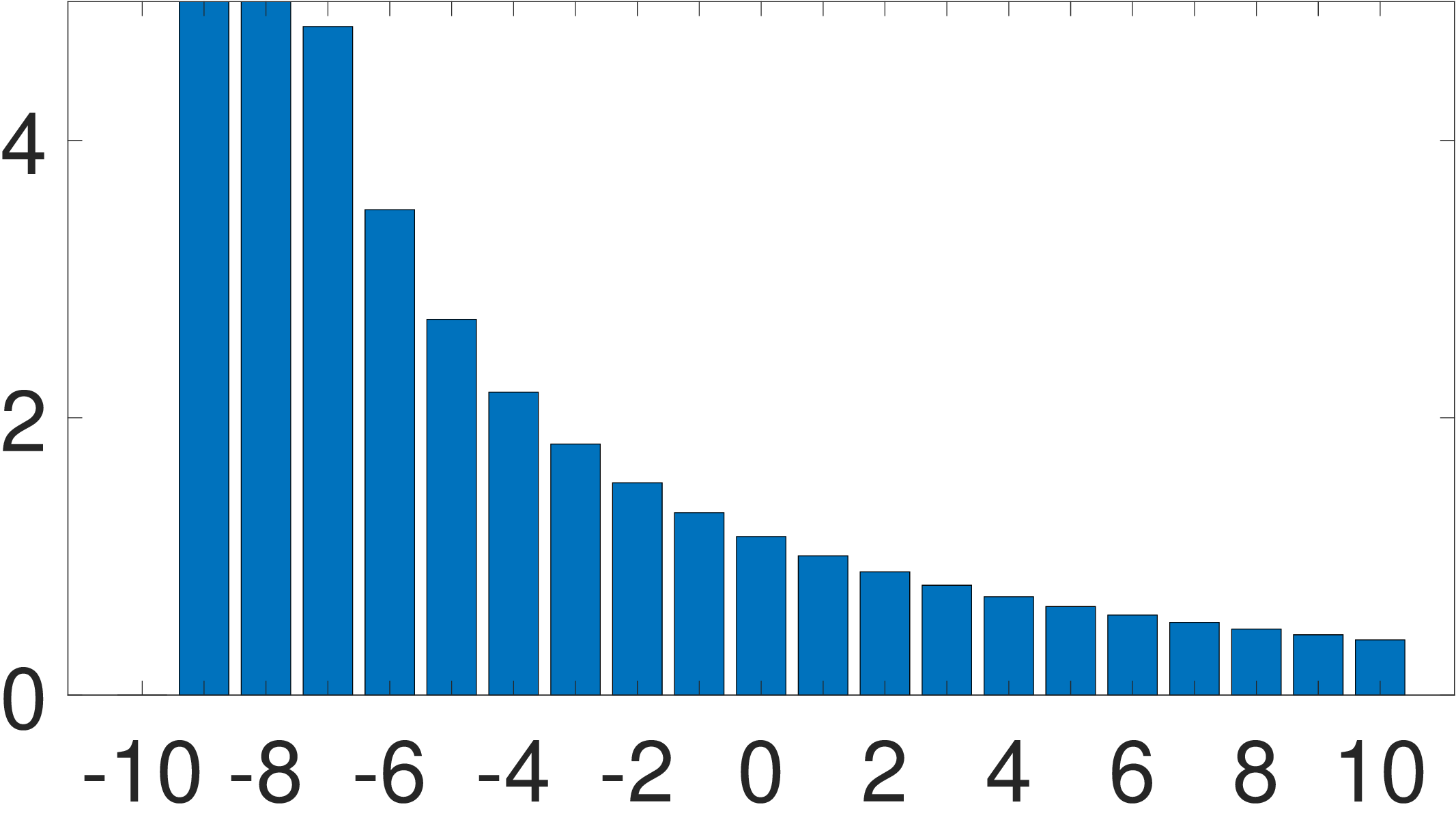}%
  \label{fig.p1p2}
}\hfill
\subfloat[][$\omega=1$.]{%
  \includegraphics[width=0.48\linewidth]{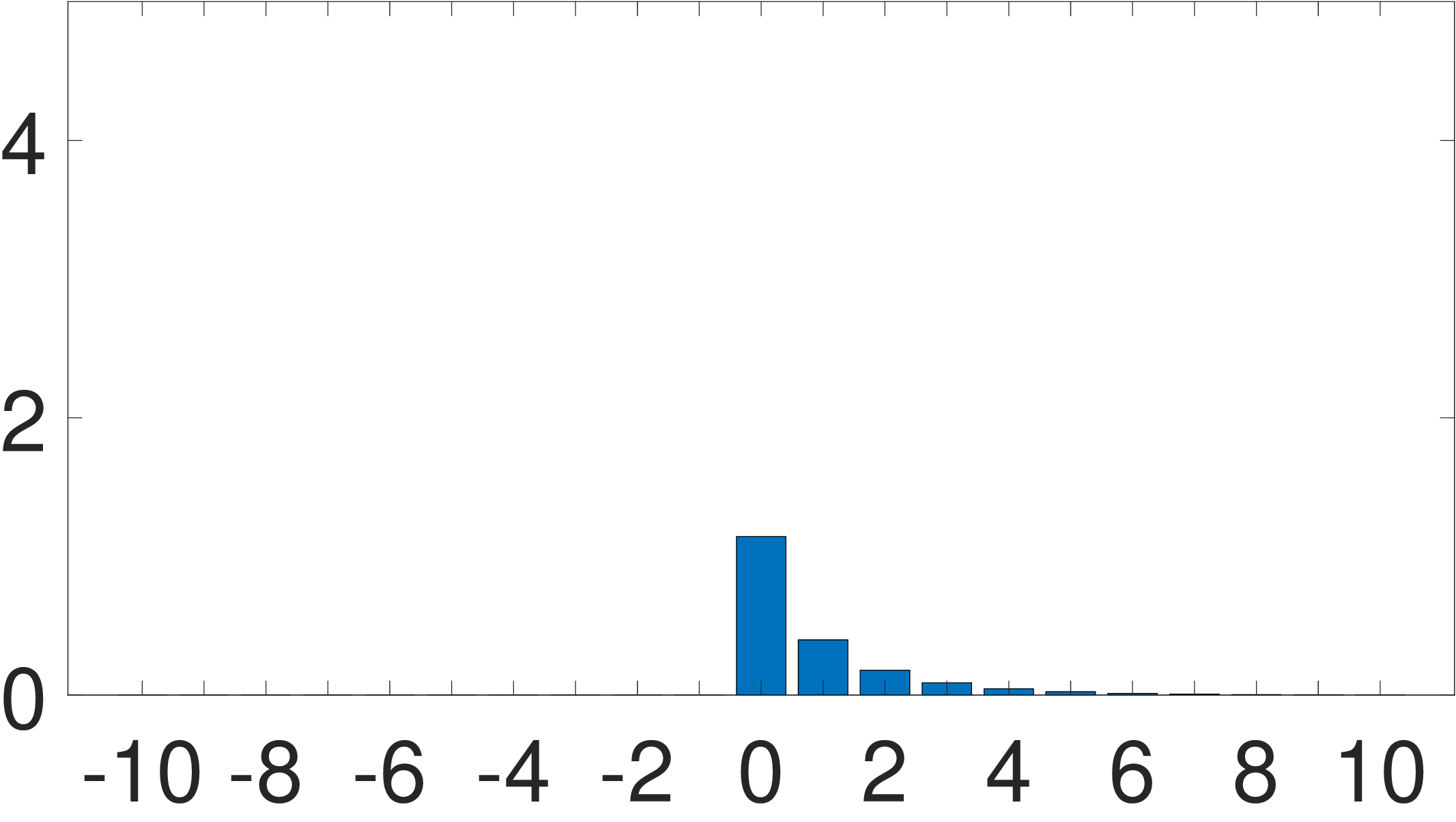}%
  \label{fig.p1p3}
}\hfill
\subfloat[][$\omega=10$.]{%
  \includegraphics[width=0.48\linewidth]{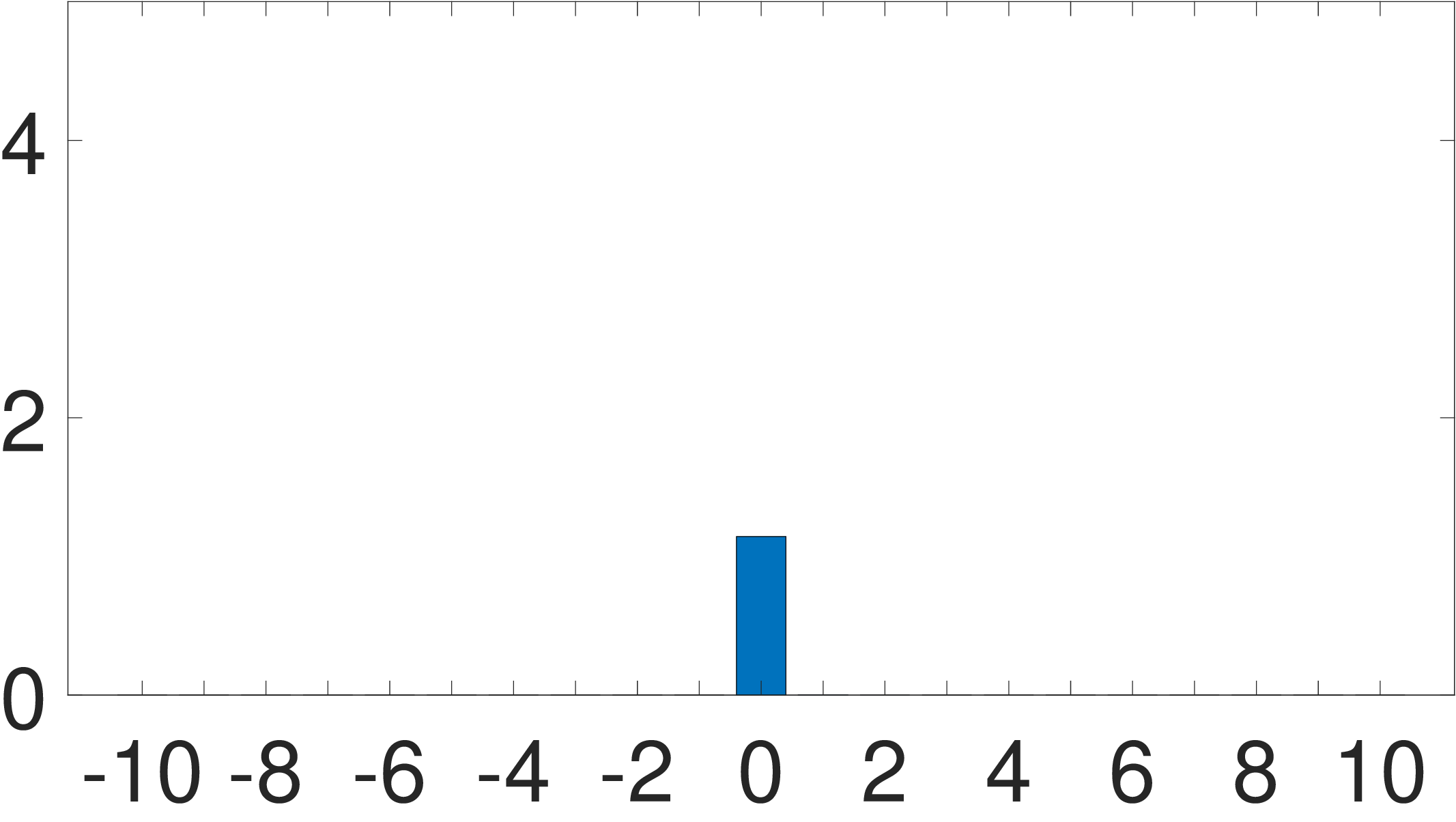}%
  \label{fig.p1p4}
}\hfill
\subfloat[][$\omega=-0.1$.]{%
  \includegraphics[width=0.48\linewidth]{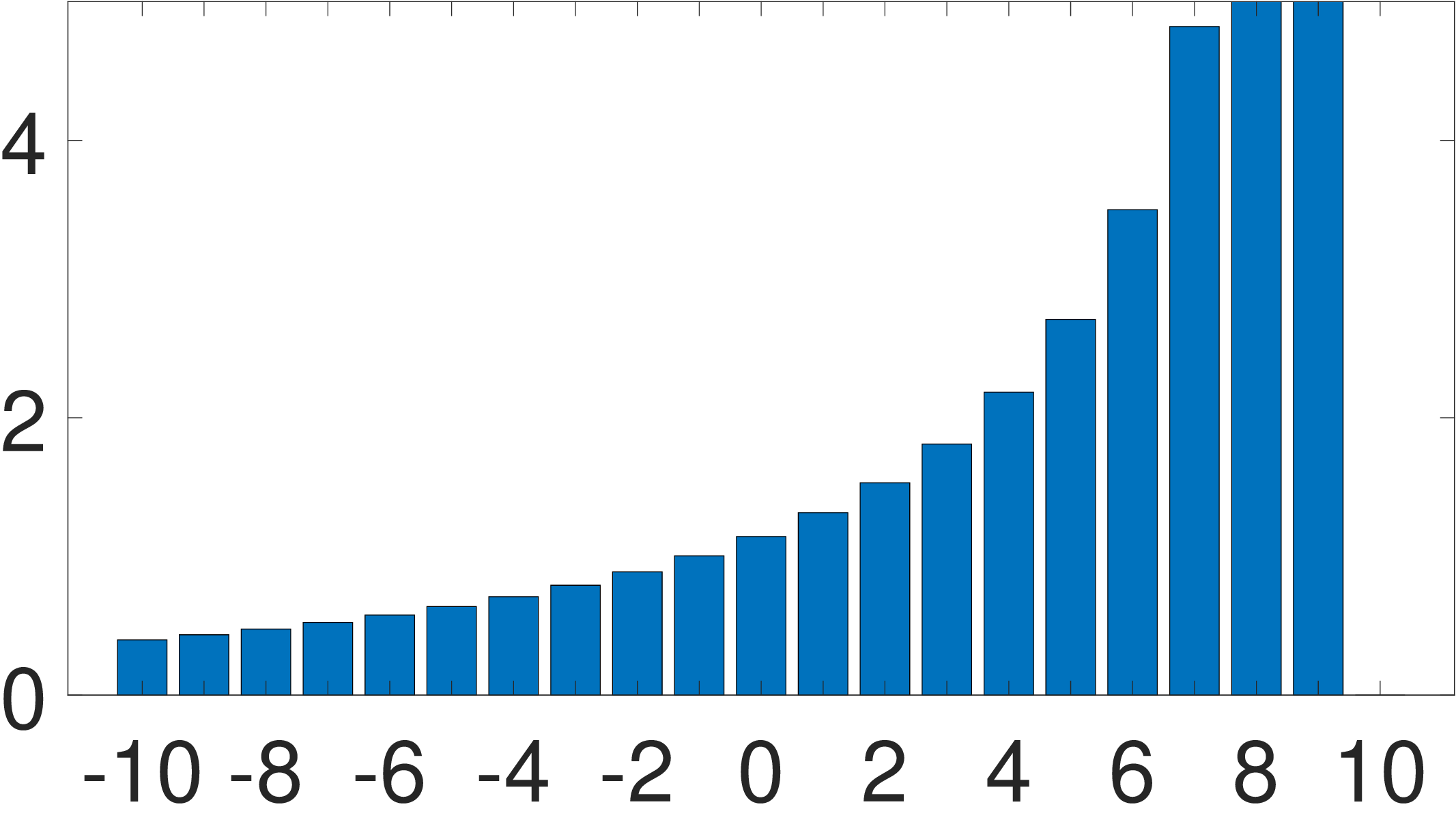}%
  \label{fig.p1p5}
}\hfill
\subfloat[][$\omega=-1$.]{%
  \includegraphics[width=0.48\linewidth]{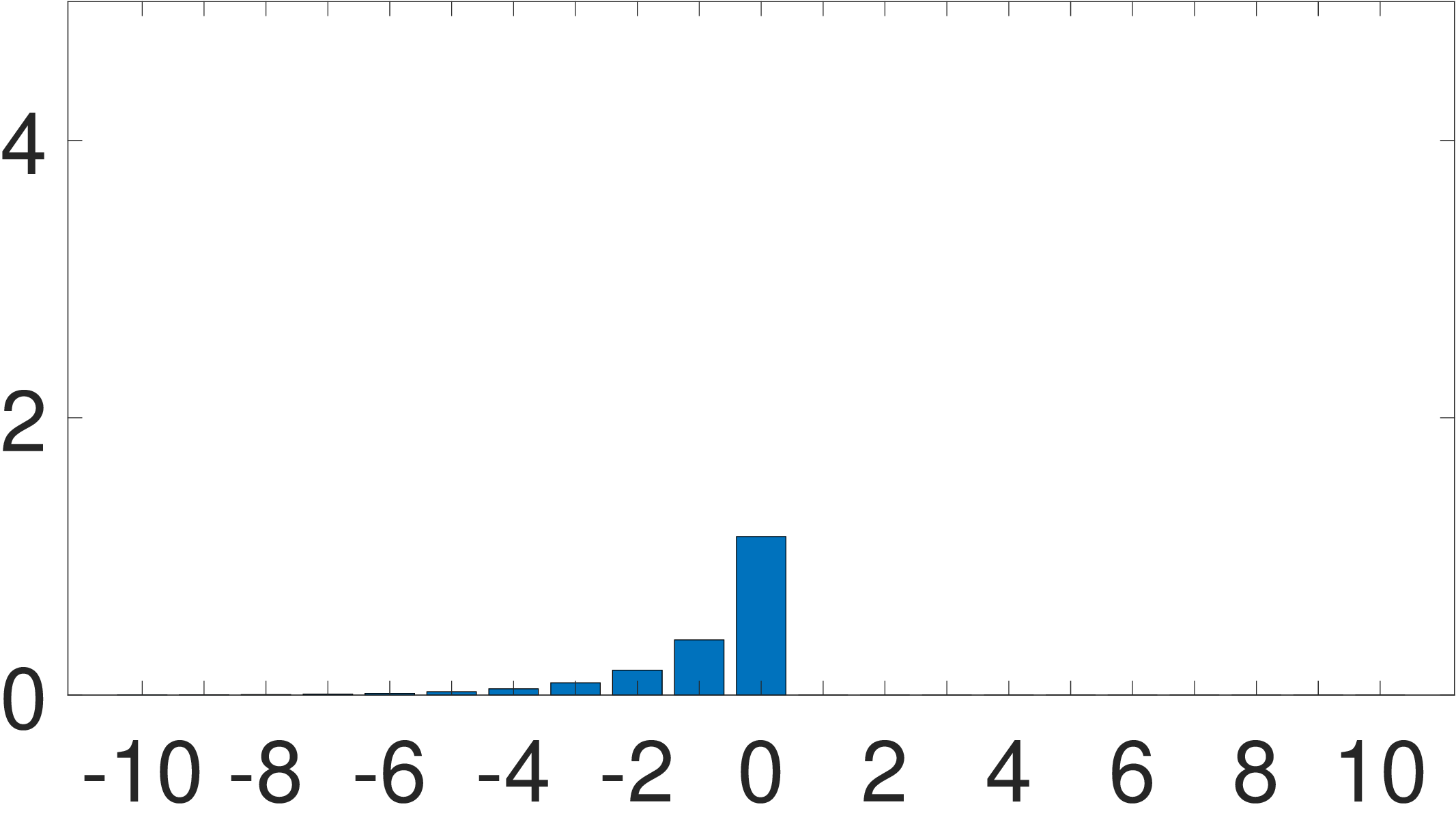}%
  \label{fig.p1p6}
}\hfill

\caption{Distributions for the expected particle number with $\mathcal E=1$ and $a=10$. The angular velocities are set to (a) 0, (b) 0.1, (c) 1, (d) 10, (e) -0.1, and (f) -1.}
\label{fig.p1p}
\end{figure}

Next, we plot the distribution for the expected particle numbers of different OAM modes. One may notice that the energy $\mathcal E$ modifies the distribution differently when it has different signs. First, suppose $\mathcal E>0$. This is the most common case, where the observer detects a particle with a definite energy $\mathcal E$ and some OAM value $l$. In Fig. \ref{fig.p1p}, we plot the particle number distributions for different angular velocities $\omega$. The acceleration $a$ is set to $10$. Figure \ref{fig.p1p1} shows that when $\omega=0$, as we find in our previous study \cite{wu2023orbital}, all OAM modes share the same particle number. The ensemble average of total OAM $\left <L_z \right >$ is zero, which reflects the fact that the spacetime structure of the transverse plane along the observer's trajectory is isotropic. When the observer starts to rotate with a positive angular velocity, the critical OAM $l_c$ moves from negative infinity to the right. As $\omega$ increases, more and more modes disappear, and the modes to the right of $l_c$ are regulated by an exponential function and form a tail, as shown in Fig. \ref{fig.p1p2}. When $\omega$ reaches $\mathcal E$, all negative OAM modes are disallowed, and the critical OAM $l_c$ stops at $\l=0$, as in Fig. \ref{fig.p1p3}. Figure \ref{fig.p1p4} shows that as $\omega$ continues to grow, the positive modes are depressed more and more strongly. If the light-speed limit is set aside for a moment, in the case where $\omega=\infty$, only the $l=0$ mode can survive. In this case, from the perspective of the observer, the surrounding spacetime is rotating with an angular velocity of $-\infty$. The spacetime points on the transverse plane will not be well defined. If any nonzero OAM shows up, it will induce an ill-defined phase factor. Hence, we expect only the zero OAM modes to survive, as our calculation shows. We also plot the probability distribution for negative angular velocity in Figs. \ref{fig.p1p5} and \ref{fig.p1p6}. They show a similar but reverse procedure as the $\omega$ gradually decreases to negative infinity.

\begin{figure} [tbhp]
\centering
\subfloat[][$\omega=0$.]{%
  \includegraphics[width=0.48\linewidth]{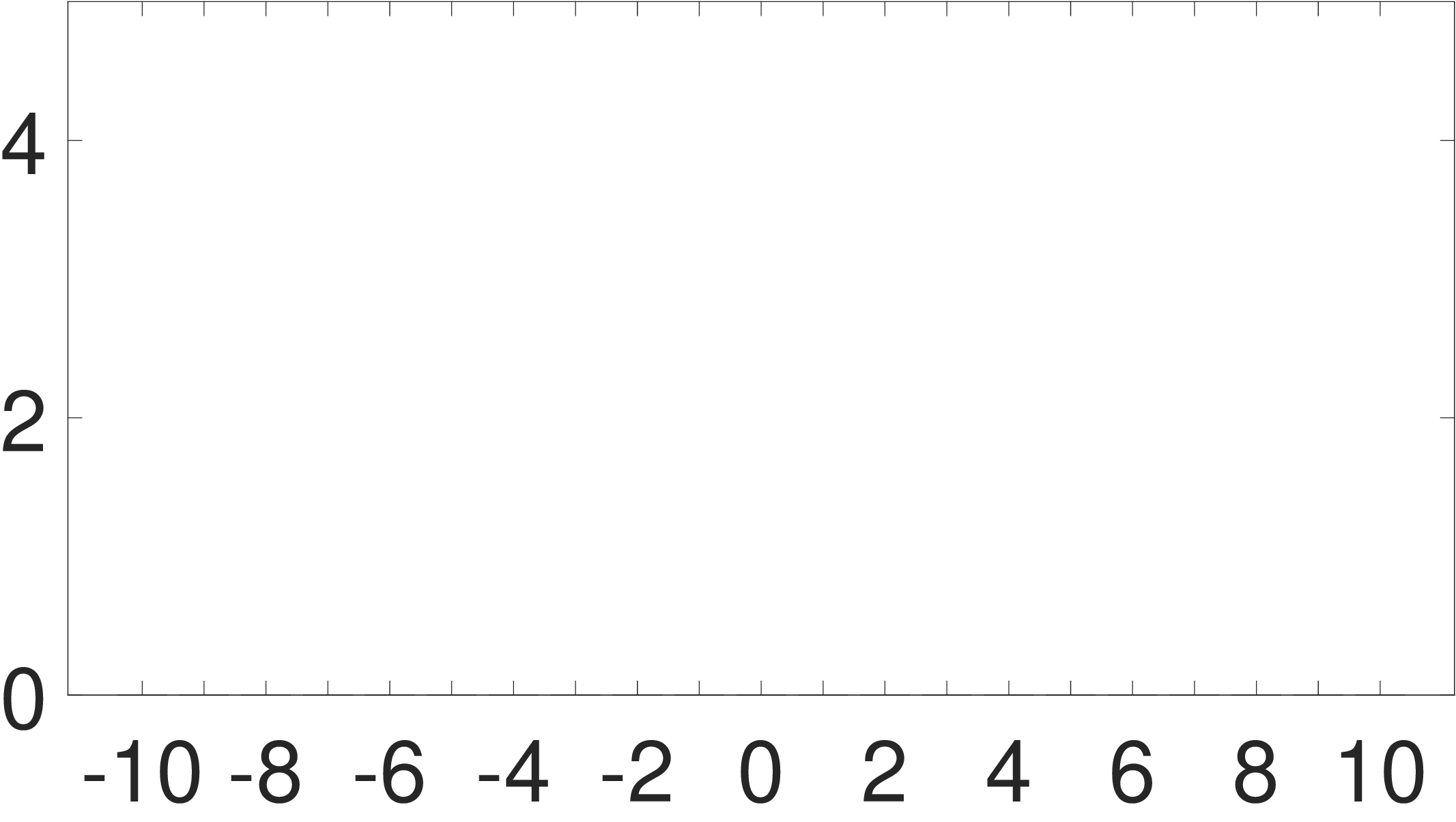}%
  \label{fig.p1m1}
}\hfill
\subfloat[][$\omega=0.5$.]{%
  \includegraphics[width=0.48\linewidth]{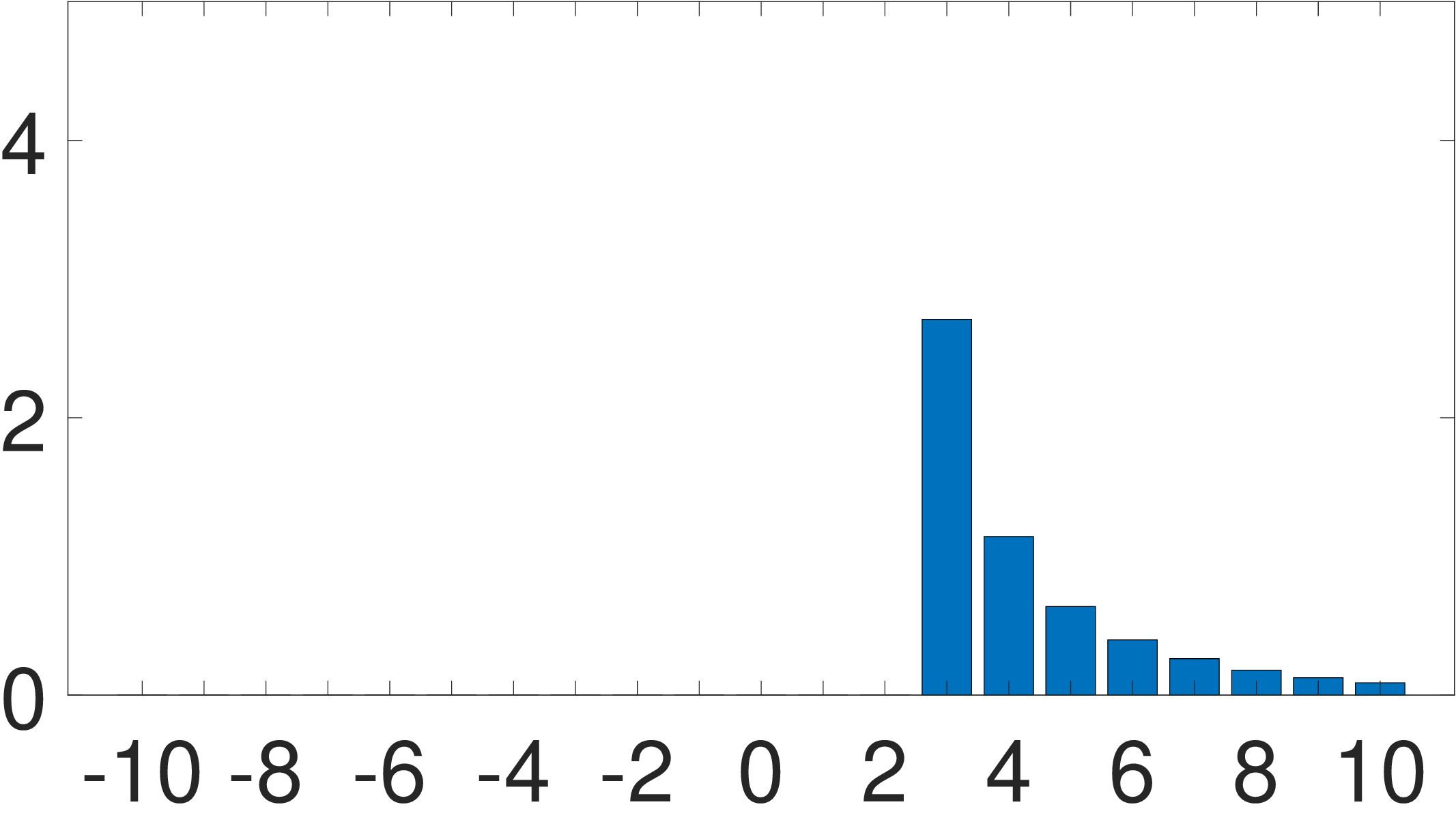}%
  \label{fig.p1m2}
}\hfill
\subfloat[][$\omega=2$.]{%
  \includegraphics[width=0.48\linewidth]{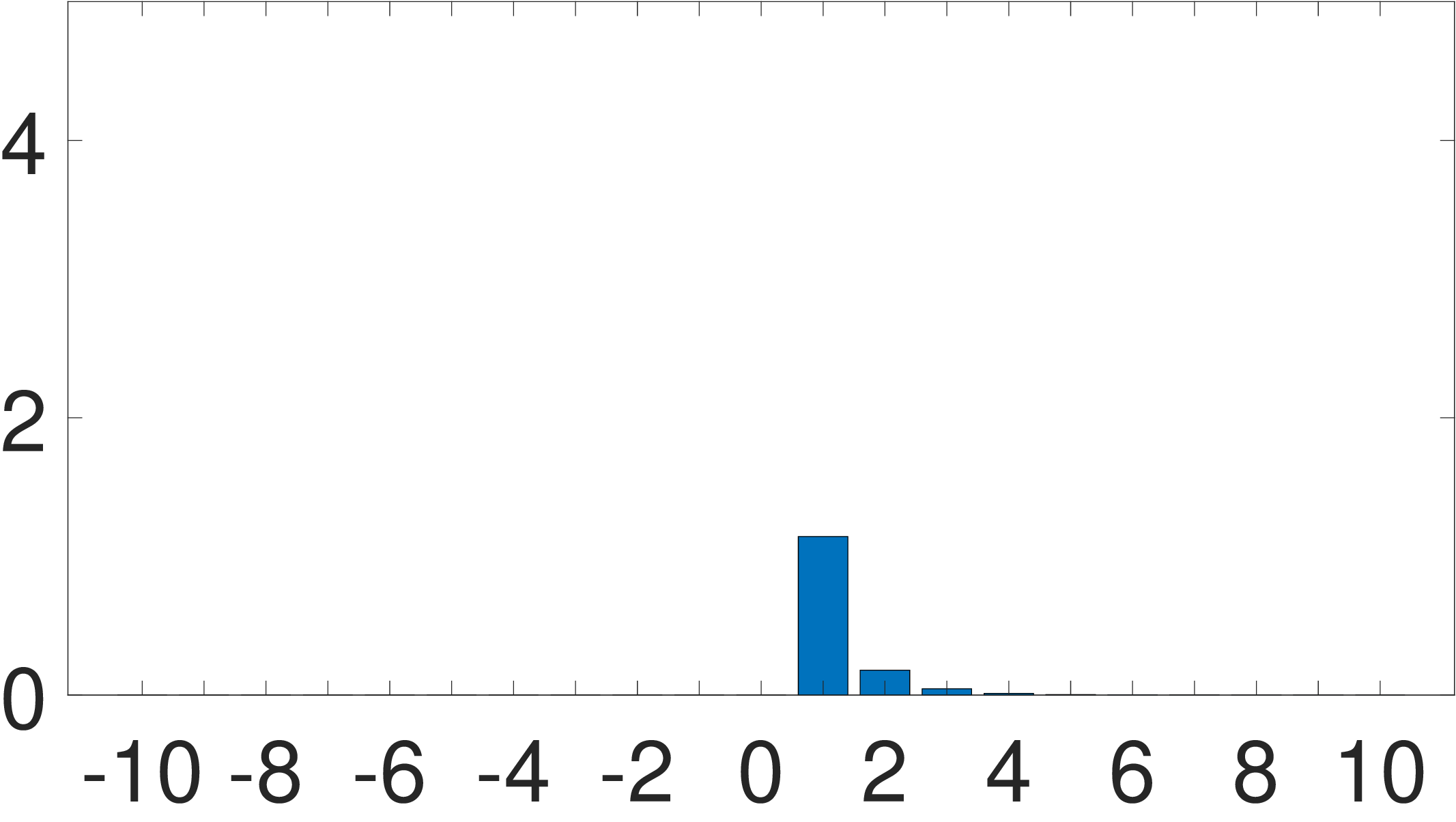}%
  \label{fig.p1m3}
}\hfill
\subfloat[][$\omega=10$.]{%
  \includegraphics[width=0.48\linewidth]{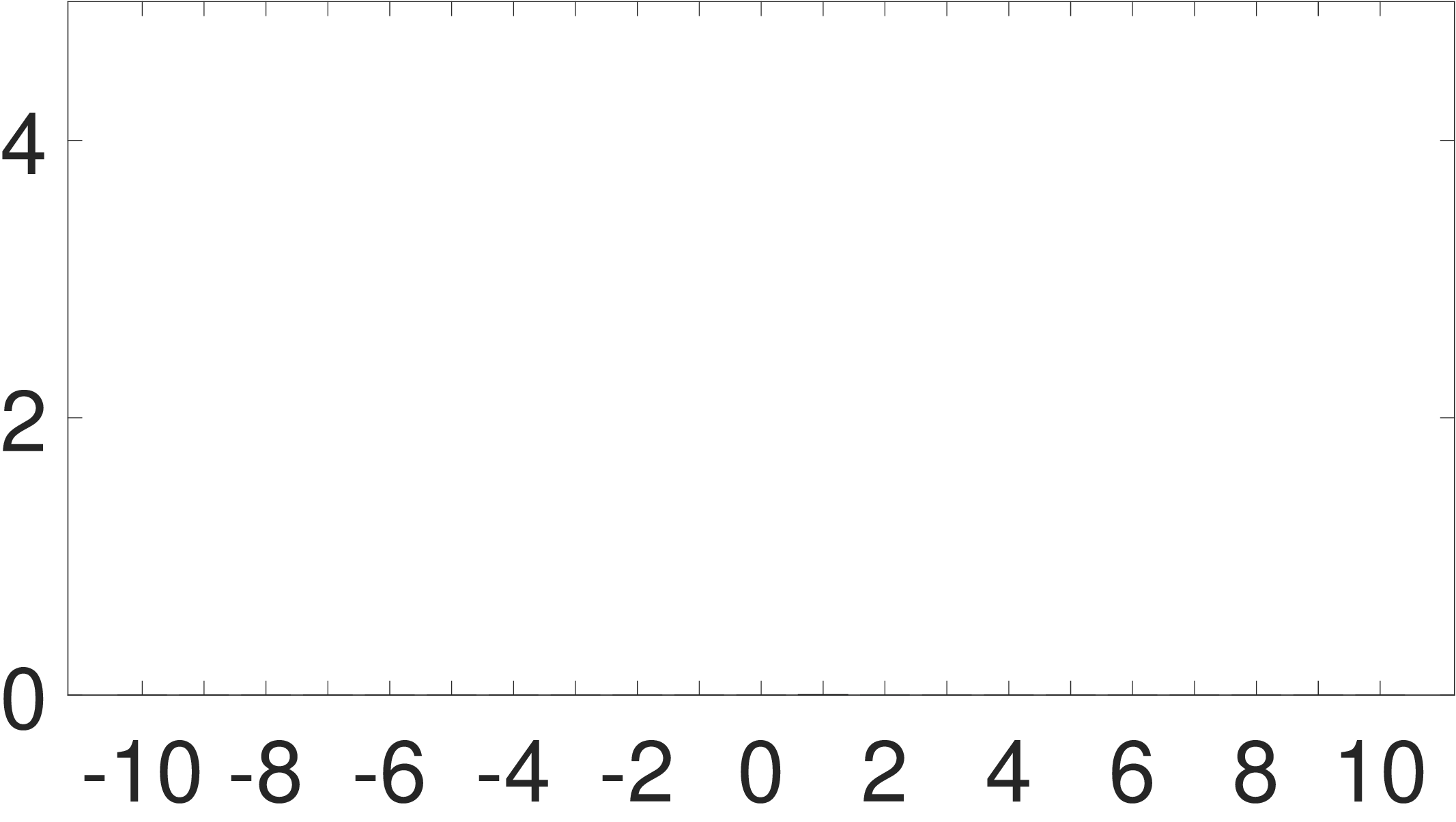}%
  \label{fig.p1m4}
}\hfill
\subfloat[][$\omega=-0.5$.]{%
  \includegraphics[width=0.48\linewidth]{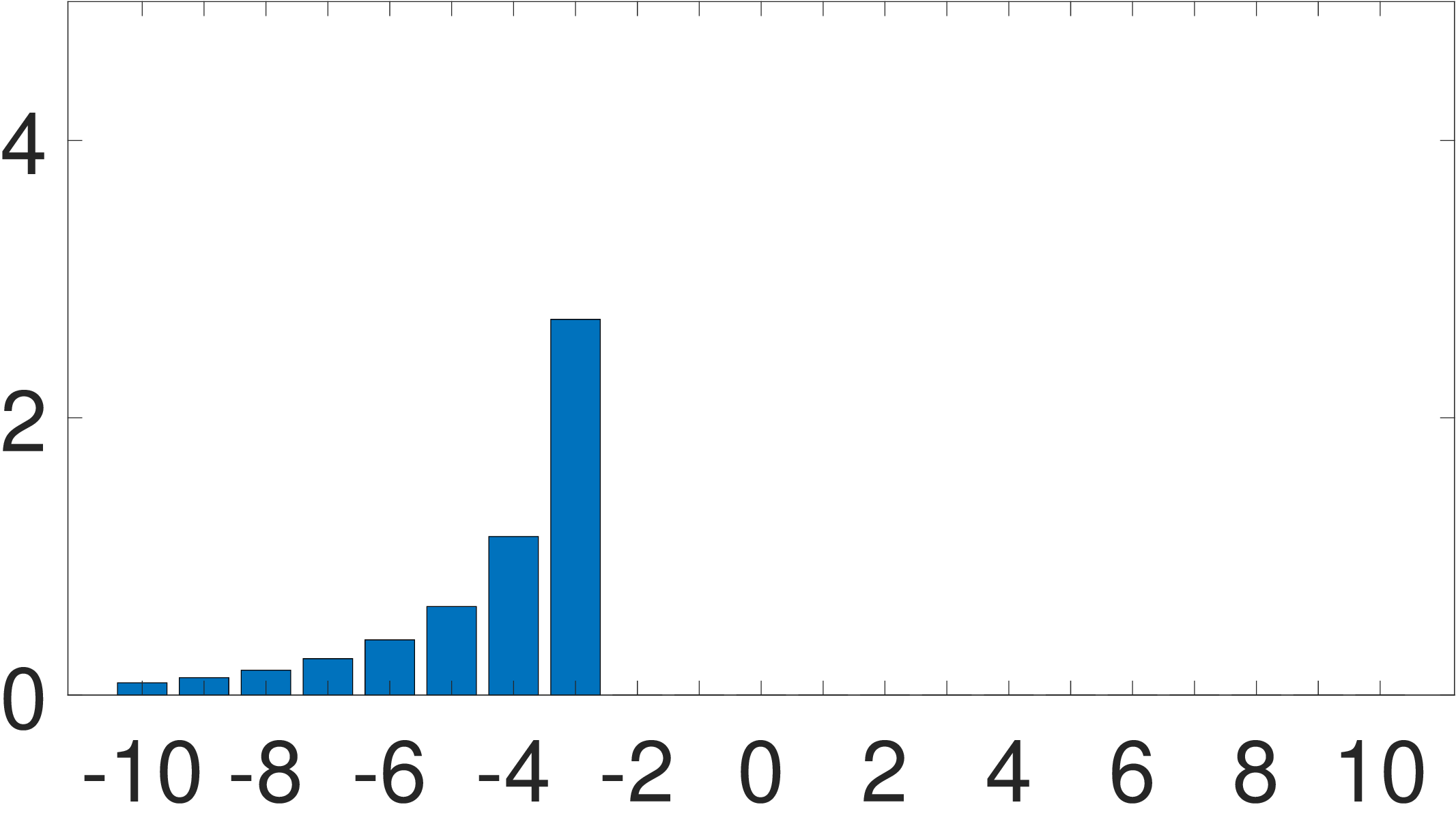}%
  \label{fig.p1m5}
}\hfill
\subfloat[][$\omega=-2$.]{%
  \includegraphics[width=0.48\linewidth]{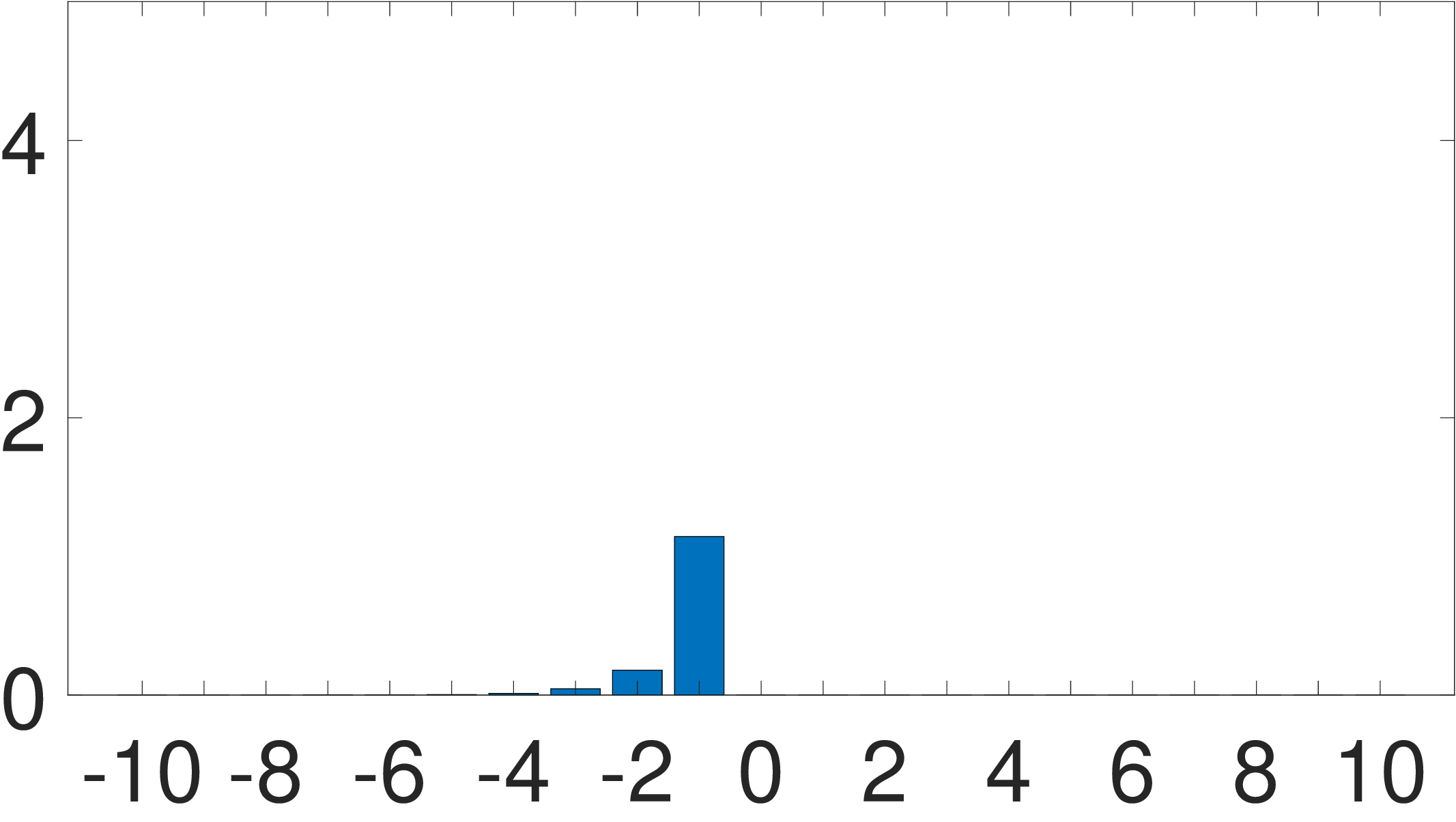}%
  \label{fig.p1m6}
}\hfill

\caption{Distributions for the expected particle number with $\mathcal E=-1$ and $a=10$. The angular velocities are set to (a) 0, (b) 0.5, (c) 2, (d) 10, (e) -0.5, and (f) -2.}
\label{fig.p1n}
\end{figure}

In contrast to the linear accelerated frame, a new kind of particle exists in the rotating case. When $\omega=0$, the particle energy $\mathcal E$ is equal to the Rindler energy $\Omega$. This requires that the particle energy always be positive. Hence, the probability of detecting a negative-energy particle is zero, as shown in Fig. \ref{fig.p1m1}. However, when the detector starts to rotate, the introduction of $l\omega$ with $l>0$ allows the existence of negative-energy modes. The detector can be excited either by absorbing a positive-energy particle or emitting a negative-energy particle. The negative particle is mainly located beyond the static limit, and this process is analogous to the Penrose effect near a rotating black hole \cite{korsbakken2004fulling}. For a given negative $\mathcal E$, the modes $l>|\mathcal E|/\omega$ can be detected, but higher-order OAM modes have a lower probability, as in Fig. \ref{fig.p1m2}. As the angular velocity grows, more and more positive modes are allowed, as in Fig. \ref{fig.p1m3}. But in the end, it will stop at $l=1$. The zero OAM and negative OAM modes are always forbidden. The expected particle number of $l\ge1$ will continue to be suppressed as $\omega$ keeps growing. When $\omega \rightarrow \infty$, all modes vanish again, as in Fig. \ref{fig.p1m4}. A similar but reverse procedure shows up when the $\omega$ is negative, as in Figs. \ref{fig.p1m5} and \ref{fig.p1m6}.

\section{Unruh-DeWitt detector for OAM modes}
In the previous section, we see that the Minkowski vacuum state can be expressed by particle states in the rotating accelerated frame, but one may ask how the observer will actually experience these states. We will use an Unruh-DeWitt detector (UD detector) to represent the observer, and we study how the detector reacts when it interacts with those states. Originally, the UD detector was designed to be a pointlike two-level monopole quantum system, which interacts with the scalar field locally via the interaction Hamiltonian \cite{unruh1976notes,birrell1984quantum,korsbakken2004fulling}, \begin{equation}
H_{I,{\rm ori}}(\tau)=c(\tau) \int d^3 {\mathbf x}' M(\tau,x^i(\tau),{x'}^j)\psi({x'}^j),\label{eq.oriInteractionHamiltonian}
\end{equation}
where $\tau$ is the proper time of the detector; $c(\tau)$ is the coupling factor, which is assumed to be small; and $M(\tau,x^i(\tau),{x'}^j)$ is the monopole moment operator of the detector. We denote the free Hamiltonians of the detector and the scalar field by $H_D$ and $H_\psi$, respectively. Since the detector is modeled to be pointlike and move along the observer's trajectory ($\tau, x^i(\tau)$) in the right region, and its interaction with the field is localized, then $M(\tau,x^i(\tau),{x'}^j)$ should have the form\begin{equation}
M(\tau,x^i(\tau),{x'}^j)=m(\tau)\delta^{(3)}({x'}^i-x^i(\tau)),
\end{equation}
where $m(\tau)$ bears the role of monopole moment operator, and the three-dimensional Dirac delta function, $\delta^{(3)}({x'}^i-x^i(\tau))$, restricts the interaction to be localized to the trajectory of the detector, i.e., $x^i(\tau)$. 

However, this UD detector may not be very useful for detecting the OAM particles because the OAM is a phase distribution over the transverse plane, and it will not yield the information for OAM by only detecting a single point. Instead, to obtain the topological charge of an OAM particle, we should extract the phase information along a transverse circle centered in the propagation axis. For this reason, we shall redesign the UD detector to be transverse-extended, i.e., we leave the longitudinal interaction to be localized. First, we assume that the detector has a ringlike shape extending from a circle with a small radius $r$, so we can extract phase information over the ring, and we write the $r'$ dependence of the monopole term as $e^{-(r'-r)^2/(2L^2)}/r'$, where $L$ is some scale parameter. The $1/r'$ factor is included simply to cancel the $r'$ factor introduced by $d^3 {\mathbf x}'=r'dr'd\theta'dx'^3$. In fact, if we leave out the $1/r'$ factor, we will only have to include a constant $r$ in the result, which can be absorbed by the weak coupling factor. This justifies the inclusion of $1/r'$. The $e^{-(r'-r)^2/(2L^2)}$ factor restricts the interaction to be near the circle. When $L$ increases, the ring extends, on the transverse plane, from the circle, and the interaction area with the field increases, as well. However, we wish to keep the interaction local, at least to some extent, as the original UD detector does, so we choose to decrease the scale parameter $L$, and the detector shrinks toward the circle. When $L$ is small enough, we could approximate $e^{-(r'-r)^2/(2L^2)}$ by $\sqrt{2\pi} L \delta(r'-r) $, where $L$ is now only a real number. Further, we would like the detector to interact only with particles with a particular OAM. Thus, let us consider a simplest case, where there are only two different states. One is the ground state with OAM 0. The other one is an excited state with OAM $l$. They will be denoted by $\left | g,0\right >$ and $\left | e,l\right >$, respectively. To let the detector couple to a particle with OAM $l$, we require that at $\tau=0$ the detector have a phase structure of $e^{\pm il\theta'}$ in the form of $\big (\left | e,l\right >\left < g,0\right |e^{-il\theta'}+\left | g,0\right >\left < e,l\right |e^{il\theta'}\big )$, where the first term describes that a detector in the ground state absorbs a particle with OAM $l$ to jump to the excited state, while the latter term describes the reverse process. There is another important modification. Note that the phase structure $e^{\pm il\theta'}$ is defined in the detector's inertial frame, or we may say that the structure is constructed before the detector starts to rotate. However, when the detector rotates, it will become $e^{\pm  il\tilde {\theta'}}$, where $\tilde {\theta'}=\theta'-\omega \tau$ is the azimuthal angle in the rest frame of the detector. Hence, when viewed from the Rindler frame, the phase structure should be written as $\big (\left | e,l\right >\left < g,0\right |e^{-il\theta'+il\omega \tau}+\left | g,0\right >\left < e,l\right |e^{il\theta'-il\omega \tau}\big )$. Later, we will see that this structure indeed ensures that the detector will only couple to particles with OAM $l$. As for the interaction along $x'^3$, it is still local and can be written as $\delta(x'^3-x^3) $. Putting these factors together, we can derive the interaction Hamiltonian for the transverse-extended UD detector to be\begin{align}
H_{I}(\tau)=&c(\tau)\int dr'd\theta'dx'^3e^{iH_D\tau}\big (\left | e,l\right >\left < g,0\right |e^{-il\theta'+il\omega \tau}+\left | g,0\right >\nonumber \\ &\times \left < e,l\right |e^{il\theta'-il\omega \tau}\big  )e^{-iH_D\tau}\sqrt{2\pi}L\delta(r'-r) \delta(x'^3-x^3)\nonumber \\&\times\psi(\tau,r',\theta',x'^3)\nonumber \\
=&c(\tau)\int_0^{2\pi} d\theta' e^{iH_D\tau}(\left | e,l\right >\left < g,0\right |e^{-il\theta'+il\omega \tau}+\left | g,0\right >\nonumber \\ &\times \left < e,l\right |e^{il\theta'-il\omega \tau})e^{-iH_D\tau}\psi(\tau,r,\theta',x^3),\label{eq.interactionHamiltonian}
\end{align}
where in the last line, $\sqrt{2\pi} L$ is absorbed into the coupling factor $c(\tau)$. For simplicity, we will assume $c(\tau)$ is invariant in time, i.e., $c(\tau)=c$. Also, in the Heisenberg picture, the interaction Hamiltonian evolves as $H_{I}(\tau)=e^{iH_D\tau}H_{I}(0)e^{-iH_D\tau}$.

Let us assume that the detector starts out in the state $\left | e_i,l_i\right >$ and interacts with a scalar field which is in the state $\left | \psi_i\right > $ at the time $\tau_i$; we can denote the state of the system as $\left |  e_i,l_i,\psi_i\right >=\left | e_i,l_i\right >\otimes\left | \psi_i\right > $. In the interaction picture, we can derive the evolution of the system to the time $\tau_f$, in the first-order approximation, as \begin{equation}
\left | \{ e_i,l_i,\psi_i\}(\tau_f)\right >=\bigg [1-i\int_{\tau_i}^{\tau_f}d\tau H_I(\tau)\bigg ] \left |   e_i,l_i,\psi_i\right >.
\end{equation}
Then the probability that the detector ends up in a different state $\left | e_f,l_f\right >$ ($e_f\ne e_i $) is given by 
\begin{align}
P_{i\rightarrow f}=&\left < e_i,l_i,\psi_i\right |\int_{\tau_i}^{\tau_f}d\tau H_I^\dagger(\tau) \left | e_f,l_f\right > \left < e_f,l_f\right |\nonumber \\
&\times\int_{\tau_i}^{\tau_f}d\tau' H_I(\tau') \left | e_i,l_i,\psi_i\right >,
\end{align}
where we have traced out the final field state $ \left | \psi_f\right >$ because we only care about how the detector reacts.

First, we consider the case that $\left |e_i,l_i\right >=\left | g,0\right > $ and $\left |e_f,l_f\right >=\left | e,l\right > $ and the energy gap between these states is $E>0$. The probability that the detector becomes excited is now given by \begin{align}
P_{g\rightarrow e}=&c^2\int_{\tau_i}^{\tau_f}d\tau\int_{\tau_i}^{\tau_f}d\tau'  e^{-iE(\tau-\tau')}e^{-il\omega(\tau-\tau')} \int_0^{2\pi}d\theta\nonumber \\&\times\int_0^{2\pi}d\theta' \left < \psi_i\right | e^{il\theta}\psi(x(\tau))e^{-il\theta'}\psi(x'(\tau'))\left |\psi_i \right >.\label{eq.pif}
\end{align}
If the initial scalar field is in the Minkowski vacuum state, i.e., $\left | \psi_i\right >=\left | 0\right >_{\rm M}$, the correlation function will be the positive-frequency Wightman function, i.e.,\begin{align}
G^+_l (x(\tau),x(\tau'))\equiv &\int_0^{2\pi}d\theta\int_0^{2\pi}d\theta' \left < 0\right |_{\rm M} e^{il\theta} \psi(x(\tau))\nonumber \\&\times e^{-il\theta'}\psi(x'(\tau'))\left |0 \right >_{\rm M}.
\end{align}
Later, it will turn out that the positive-frequency Wightman function only depends on the proper time difference $\tau-\tau'$, so we may write it as $G^+_l (\tau-\tau')$. By changing variables from $\tau$ and $\tau'$ to $u=\tau+\tau'$ and $s=\tau-\tau'$, we can derive the transition rate as\begin{equation}
\Gamma_{g\rightarrow e}=2c^2\int_{\tau_i-\tau_f}^{\tau_f-\tau_i} e^{-iEs-il\omega s} G^+_l(s)ds.\label{eq.gamma}
\end{equation}
To avoid any transient effects and to let the detector reach a thermal equilibrium with the scalar field, we can further take the limits $\tau_i \rightarrow -\infty$ and $\tau_f \rightarrow \infty$.

To calculate the positive-frequency Wightman function $G^+_l (x,x')$, we use the $(\chi,\xi,r,\theta)$ coordinates and expand the Minkowski vacuum state by Eq. \eqref{eq.MinkowskiVacuum}. When expanding the field operator $\psi(x)$ by Eq. \eqref{eq.psiRindler}, we only keep the modes with $\sigma=+$. This is because when we calculate the interaction Hamiltonian, Eq. \eqref{eq.interactionHamiltonian}, the Dirac delta function, $\delta(x'^3-x^3)$, restricts the field operator to the points followed by the detector. Since the detector is running in the right region, the modes with $\sigma=-$, which have no definition for $\xi>0$, will not take part in the calculation. Also, the points along the detector's trajectory are given by Eqs. \eqref{eq.worldline1} and \eqref{eq.worldline2}. One may verify that $G^+_l (x,x')$ indeed only depends on $s=\tau-\tau'$. The result is\begin{align}
G^+_l (s)=&\sum_m\sum_Q \int_0^\infty d\Omega \left [\frac {J_m(\tilde u_{m,Q}r_0)K_{i\Omega}(\tilde u_{m,Q}/a)}{r_*J_{m+1}(u_{m,Q})(1-e^{-2\Omega \pi})}\right ]^2  \nonumber \\&\times  \frac {4  \sinh(\Omega \pi)}{\pi}  \bigg [e^{-2\Omega \pi}e^{i\Omega a s}\delta(l-m) +e^{-i\Omega as} \nonumber \\
&\times \delta(l+m)\bigg ].
\end{align}
Note that the Rindler energy is related to the proper energy by Eq. \eqref{eq.energy}. By substituting $G^+_l (s)$ into Eq. \eqref{eq.gamma}, we can derive the transition rate to be
\begin{align}
&\Gamma_{g\rightarrow e}\nonumber \\ =&C\sum_m\sum_Q \int_{-m\omega}^{\infty} d\mathcal{E} 
\left [\frac {J_m(\tilde u_{m,Q}r_0)K_{i(\mathcal E+m\omega)/a}(\tilde u_{m,Q}/a)}{J_{m+1}(u_{m,Q})(1-e^{-2\pi(\mathcal E+m\omega)/a })}\right ]^2 \nonumber \\&\times \sinh\left (\frac {(\mathcal E +m\omega)\pi}{a} \right ) \Big [e^{-2\pi (\mathcal E+m\omega)/a}\delta(l-m) \delta(E-\mathcal E)\nonumber \\ &+\delta(l+m)\delta(E+\mathcal E) \Big ],
\end{align}
where all unimportant constants are absorbed into a single constant $C$. The terms in the second square bracket bear the selection rules that we are looking for. The first term describes the process in which the detector absorbs a particle in state $\left | E, l\right > $ to jump from the ground state to the excited state, while the second term says that the detector emits a particle in state $\left |- E, -l\right > $, or equivalently, the detector absorbs a particle in state $\left | E, l\right > $. Hence, the detector is only coupled to the particles with OAM $l$, as we required. The summation related to $Q$ is messy, but it will turn out to be insignificant, so we just write it as $I(\mathcal E,m,\omega)=\sum_{Q=1}^\infty   [J_m(\tilde u_{m,Q}r_0)K_{i(\mathcal E+m\omega)/a}(\tilde u_{m,Q}/a)/J_{m+1}(u_{m,Q})]^2$ with a useful property that $I(\mathcal E,m,\omega)=I(-\mathcal E,-m,\omega)$. Then, the transition rate is given by \begin{align}
&\Gamma_{g\rightarrow e}\nonumber \\ =&C \int_{-l\omega}^\infty d\mathcal E \frac {\sinh((\mathcal E+l\omega)\pi/a)}{1-e^{-2 (\mathcal E+l\omega)\pi/a}} I(\mathcal E,l,\omega) e^{-2 (\mathcal E+l\omega)\pi/a}\nonumber \\ &\times\delta(E-\mathcal E)+C \int_{l\omega}^\infty d\mathcal E  \frac {\sinh((\mathcal E-l\omega)\pi/a)}{1-e^{-2 (\mathcal E-l\omega)\pi/a}}\nonumber \\ &\times I(\mathcal E,-l,\omega)\delta(E+\mathcal E).
\end{align}
Note that the lower limits of the integral, $\pm l\omega$, are excluded. Similarly, by letting $\left |e_i,l_i\right >=\left | e,l\right > $ and $\left |e_f,l_f\right >=\left | g,0\right > $, one can derive the transition rate, $\Gamma_{e\rightarrow g}$, from the excited state to the ground state .

\begin{figure} [tbhp]
\centering
\subfloat[][$E\le -l\omega $]{%
  \includegraphics[width=0.25\linewidth]{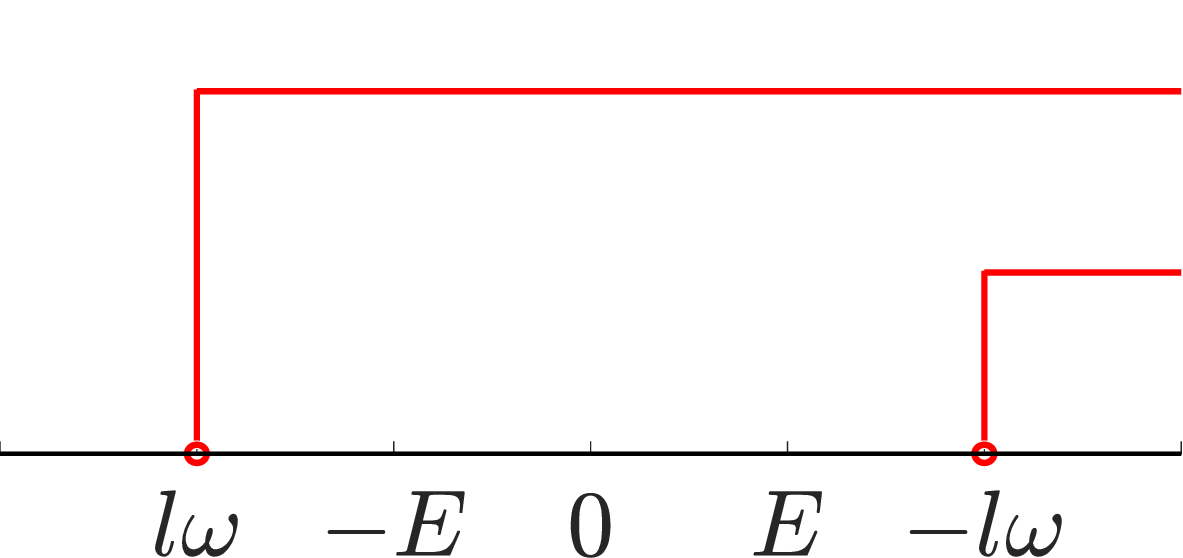}%
  \label{fig.axis1}
}\hfill
\subfloat[][$E\le l\omega $]{%
  \includegraphics[width=0.25\linewidth]{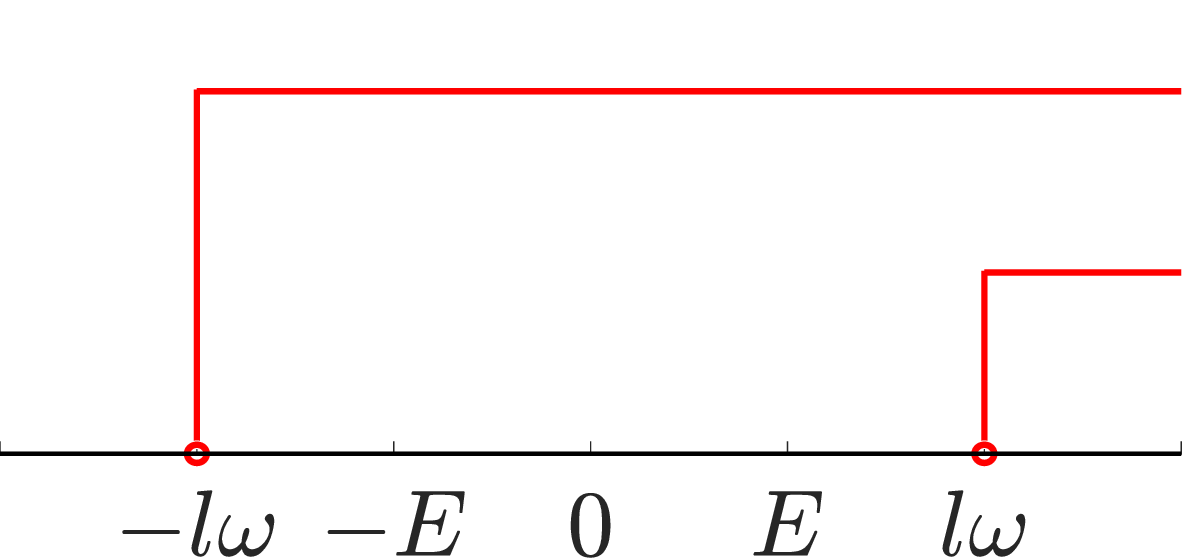}%
  \label{fig.axis2}
}\hfill
\subfloat[][$E>\left | l\omega \right |$]{%
  \includegraphics[width=0.25\linewidth]{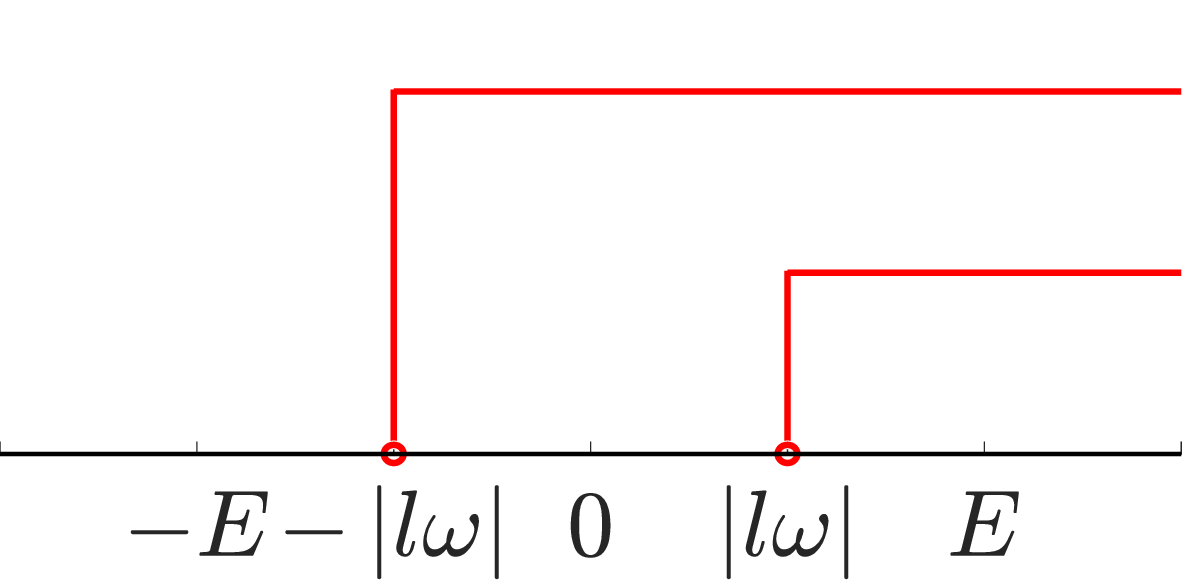}%
  \label{fig.axis3}
}\hfill
\caption{The integral ranges for three different cases. The red lines indicate the integral ranges for the two integrals in the transition rates. Note that the points of $\pm l\omega$ are excluded from the integrals.}
\label{fig.axis}
\end{figure}

\begin{figure} [tbhp]
	\centering
	\includegraphics[width=0.9\linewidth]{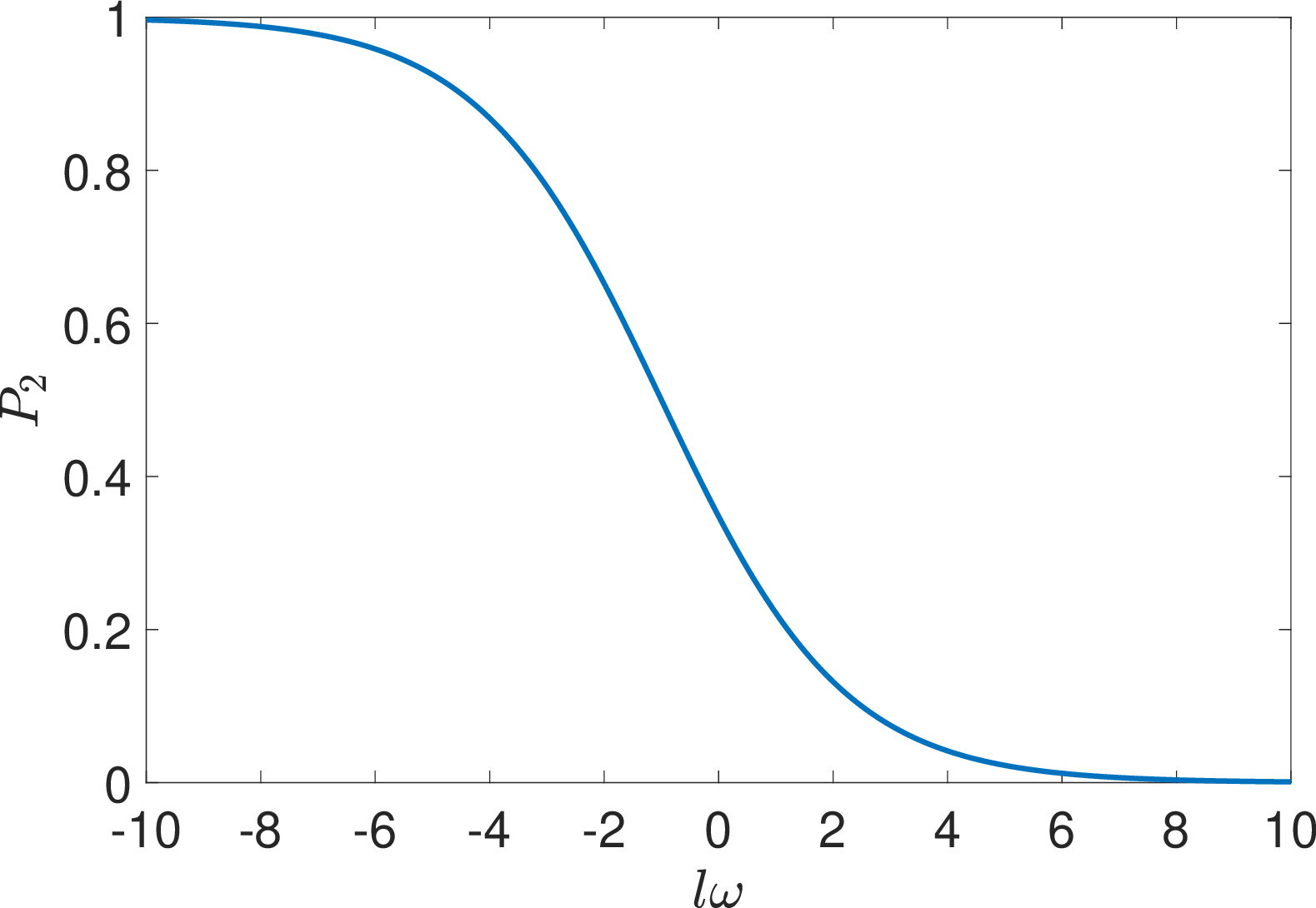}
	\caption{Probability that the detector is in the states $\left | e,l\right >$. $E$ and $a$ are set to 1 and 10, respectively.}
	\label{fig.P2}
\end{figure}

The detailed balance relation states that if the detector is initially in thermal equilibrium with the field, the populations of different states will not change in time after the detector starts to interact with the field, i.e.,
\begin{equation}
\frac {P_e}{P_g}=\frac {\Gamma_{g\rightarrow e}}{\Gamma_{e\rightarrow g}},
\end{equation}
where $P_g$ and $P_e$ are the probabilities that the detector is in the states $\left | g,0\right >$ and $\left | e,l\right >$, respectively. This relation will show us the thermal characteristics of Rindler particles. If $\omega=0$ or $l=0$, one may verify that the detailed balance is given by \begin{equation}
\frac {P_e}{P_g}=e^{-2\pi E/a}.
\end{equation}
This is precisely the Boltzmann distribution for the energy gap $E$ in the Davies-Unruh temperature $T=a/(2\pi)$. If the detector starts to rotate, the transition rates should be calculated in three different cases based on the relation between $E$ and $\pm l\omega$, and these cases correspond to different processes. We plot these cases in Fig. \ref{fig.axis}, where the red lines indicate the integral ranges for the two integrals in the transition rates. In the first case, where $E\le -l\omega$, only integrals with the lower limit being $l\omega$ survive. Hence, $\Gamma_{g\rightarrow e}$ corresponds to the process where the detector emits a particle in $\left | -E,-l\right >$, and for $\Gamma_{e\rightarrow g}$, the detector will absorb one such particle. In contrast, when $E\le l\omega$, only integrals with the lower limit being $-l\omega$ contribute, and the transition rates $\Gamma_{g\rightarrow e}$ and $\Gamma_{e\rightarrow g}$ are related to the absorbing or emitting particles in $\left | E,l\right >$ by the detector, respectively. In the last case, where $E>\left | l\omega\right |$, only terms with $\delta(\mathcal E-E)$ are nonzero, and the transition rates describe the same process as in the second case. However, all cases will yield the same detailed balance relation,\begin{equation}
\frac {P_e}{P_g}=e^{-2\pi (E+l\omega)/a}.\label{eq.P21}
\end{equation}
This relation can be interpreted in two different kinds of frames. In the eye of an inertial observer, this is still a Boltzmann distribution, but the energy needed to jump from $\left | g,0\right >$ to $\left | e,l\right >$ is shifted by $l\omega$. One may ask the reason for this energy shift. First, we note that the energy of a Rindler particle is defined in the rest frame of the detector, as stated before Eq. \eqref{eq.definitionOfE}. Meanwhile, the energy registered by an inertial observer should be calculated in the comoving inertial frame of the detector. This frame is different from the rest frame, since the latter one can be noninertial. Let us denote the coordinates in the comoving inertial frame as $(t,r,\theta,z)$. The tangent vectors of the observer's worldline form the Killing vector field in Eq. \eqref{eq.KillingField}. The last term, $a\mathbf{K_z}$, is only responsible for the generation of Rindler particles, so we can focus on the rest part, which defines the rotational motion of the detector on the transverse plane, and redefine the Killing vector field as $\mathbf v=\partial_{t}+\omega \partial_{\theta}$. We can transform to a rotating coordinate system $(\tau,r',\theta')$ by \begin{equation}
\tau=t,~~r'=r,~~\theta'=\theta-\omega t,
\end{equation}if this system is adapted to the detector by requiring $\tau$ to be its proper time \cite{letaw1980quantized}. The rotating coordinate system is the one that is defined in the rest frame. Suppose the detector can be excited by a particle with energy $E$ and OAM $l$ when it does not rotate, so its transition probability contains factors of $e^{-iE t}e^{il\theta}$ in the comoving inertial frame, as can seen from Eq. \eqref{eq.pif} with $\omega=0$. Then, after the detector starts to rotate, $t$ and $\theta$ should be replaced by $\tau$ and $\theta'$. Hence, these factors will be transformed back to the inertial frame as $e^{-i(E+l\omega) t}e^{il\theta}$. Therefore, the energy registered by an inertial observer is shifted by $l\omega$. We plot the population of the excited state $\left | e,l\right >$ in Fig. \ref{fig.P2} with $E$ being set to 1. When $l\omega$ increases, the energy gap from $\left | g,0\right >$ to $\left | e,l\right >$ becomes greater, resulting in a higher probability that the detector stays in the state $\left | g,0\right >$. On the other hand, when $l\omega$ decreases, the energy gap is narrowed down, and it is easier for the detector to jump to the state $\left | e,l\right >$. When $l\omega$ approaches $-E$ from the right, the energy gap will approach zero. The energy-time uncertainty principle requires that the transition time approach infinity. This makes the detector unable to determine whether it detects a particle or not. If $l\omega$ keeps decreasing, the two detector states will exchange their roles. $\left | e,l\right >$ will have a lower energy eigenvalue than $\left | g,0\right >$. In this case, the detector actually emits a positive energy particle to jump to $\left | e,l\right >$. As $l\omega$ goes lower, $\left | e,l\right >$ will have lower and lower energy eigenvalues, and it would be easier for the detector to decay to $\left | e,l\right >$.

It will be a different story in the rest frame, where the detector always interacts with particles in states $\left | E,l\right >$ when $l\omega>-E$. Let us start from $l\omega=0$, where, as usual, the detector detects a thermal Rindler particle bath. When $l\omega$ increases, the average particle number in the field is diminished, and it is harder and harder for the detector to jump to a higher energy state. A reverse process happens when $l\omega$ decreases. When $l\omega$ approaches $-E$ from the right, the energy gap in the inertial frame will approach zero, while the average particle number explodes to infinity. It now has equal rates for absorption and stimulated emission. Also, these particles are vacuum noise, which would impair the detection of entanglement when we explore spatial entanglement later. This would cause the phenomena of decoherence and entanglement degradation. When $l\omega$ reaches $-E$, the mode with energy $E$ is forbidden. Instead, the negative-energy mode takes its place. Now, if the detector wants to jump to $\left | e,l\right >$, it cannot absorb a $\left | E,l\right >$ particle, since there are none of them; it actually emits a $\left | -E,-l\right >$ particle. If $l\omega$ keeps decreasing, the average particle number of $\left | -E,-l\right >$ in the field decreases, as well. Therefore, the detector has a higher chance to emit such a particle and jump to state $\left | e,l\right >$.

\section{OAM entanglement in rotating accelerated frame}
Before we explore the OAM entanglement, we shall study the one-particle state in Minkowski spacetime. In previous sections, we define a new set of Minkowski modes $f^{(\sigma)}_{m\Omega Q}(x^\mu)$ by their annihilation and creation operators in Eqs. \eqref{eq.aNew} and \eqref{eq.adaggerNew}. But what kind of particles do these modes describe? Clearly, by the definition of the function $P^{(\sigma)}_{\Omega}(k_3)$, their labels, and the single-mode Bogoliubov transformation, we know they define particles with OAM $m$, Rindler energy $\Omega$, and transverse wave number $Q$. These particles are no longer monochromatic wave packets; instead, they are linear combinations with a distribution $P^{(\sigma)}_{\Omega}(k_3)$ of these wave packets. There is another important question: How will these particles move? More specifically, can these particles be detected by an observer in the Rindler right region or in the Rindler left region? The original modes $g_{lEk_3}(x^\mu)$ describe particles that can be detected either in the Rindler right region or in the Rindler left region. Nothing stops us from doing this. Just by switching the sign of $k_3$, the particles will move in the opposite region. Will the particles defined by $f^{(\sigma)}_{m\Omega Q}(x^\mu)$ behave similarly? We start by letting the operator $a^{(+)\dagger}_l$ act on the Minkowski vacuum state, \begin{equation}
a^{(+)\dagger}_l \left | 0\right >_{\rm M}=\left | 1_l\right >_{\rm M},
\end{equation}
where we omit the labels $\Omega$ and $Q$, since we mainly study the degrees of freedom of OAM. By using the single-mode Bogoliubov transformation, Eq. \eqref{eq.singleModeTransformation}, we have\begin{align}
\left | 1_l\right >_{\rm M}=&Ce^{-(\mathcal E+l\omega)\pi/(2a)}\sqrt{2\sinh{\frac {(\mathcal E+l\omega)\pi}{a}}}\sum_{\{n_m=0\}}^\infty\nonumber \\ &\times(-1)^{\sum_j jn_j} e^{-\pi\sum_j (\mathcal E +j\omega)n_j/a}\sqrt{n_l+1}\nonumber \\ &\times\left |\cdots,n_m,\cdots,n_l+1,\cdots \right >_{\rm R} \left |\cdots,n_m,\cdots,n_l,\cdots \right >_{\rm L}.
\end{align}
From this expression, we can see that the operator $a^{(+)\dagger}_l$ defines a particle moving in the Rindler right region, i.e., it can only be detected by an observer moving in the Rindler right region. Similarly, $a^{(-)\dagger}_l$ will create a particle moving in the opposite region.

\begin{figure} [tbhp]
\centering
\subfloat[][Purities.]{%
  \includegraphics[width=0.9\linewidth]{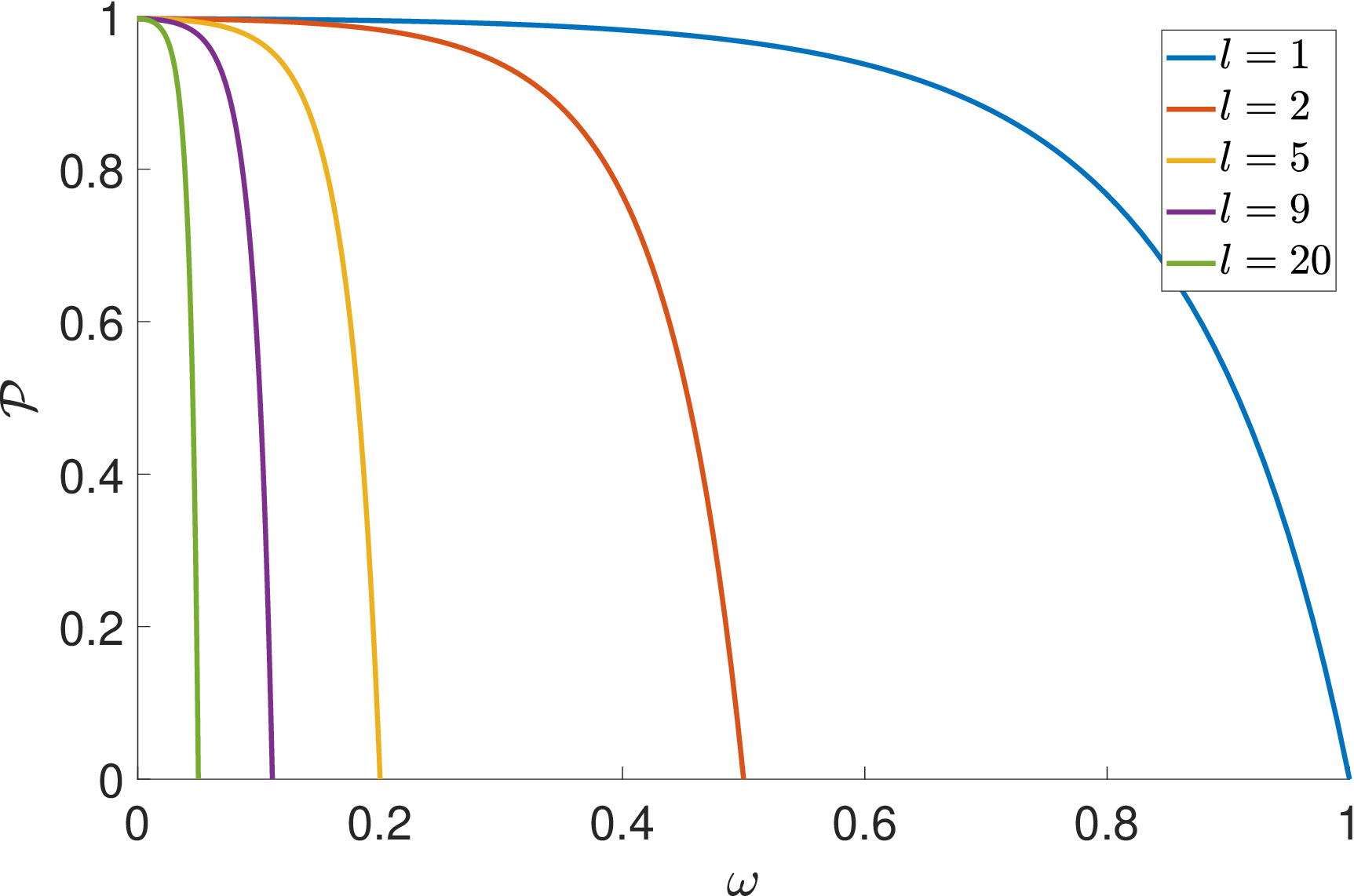}%
  \label{fig.purity2D}
}\hfill
\subfloat[][Negativities.]{%
  \includegraphics[width=0.9\linewidth]{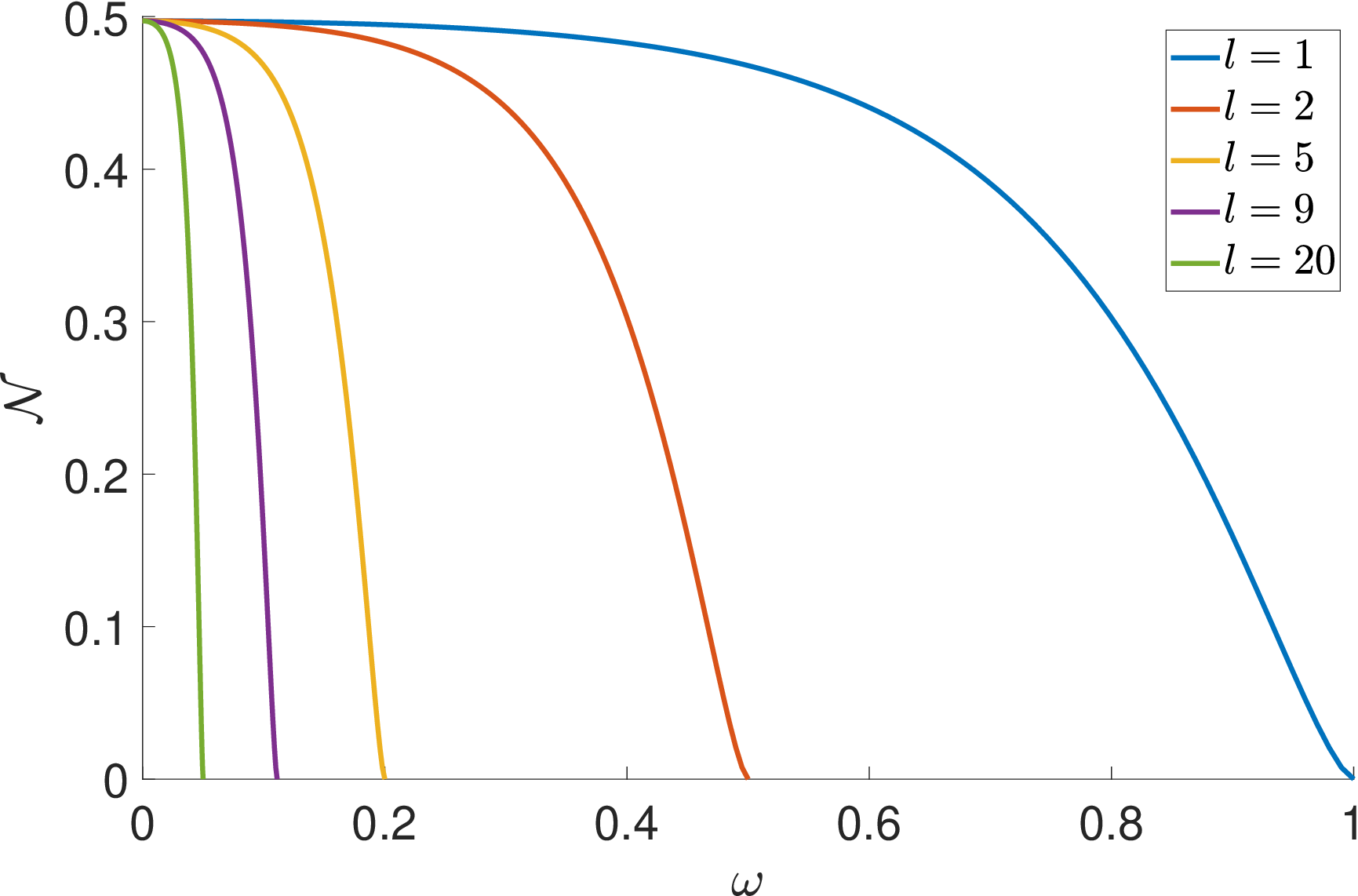}%
  \label{fig.negativity2D}
}\hfill
\caption{Curves for (a) purities and (b) negativities with OAM $l=$1, 2, 5, 9, and 20. We set $\mathcal E=1$ and $a=1$.}
\label{fig.entanglement2D}
\end{figure}

Now we study the OAM entanglement based on the one-particle states. Unlike the degrees of freedom of spin angular momentum (SAM), which is specified in a finite-dimensional space, e.g., 2 for photons, OAM can assume well-defined values of $l \hbar $ where $l=0, \pm 1, \pm 2, \dots$ that span an infinite-dimensional Hilbert space \cite{allen1992orbital,calvo2006quantum,mair2001entanglement,molina2001management,leach2002measuring,vaziri2002experimental,vaziri2003concentration,allen1999iv,kivshar2001optical,molina2007twisted,torres2011twisted,yao2011orbital,molina2001propagation}. Due to the fact that information may be encoded in a high-dimensional space, OAM is a potential source for future quantum communications \cite{vaziri2003concentration,mair2001entanglement,vaziri2002experimental}. For now, let us concentrate on the two-dimensional entanglement, defined by \begin{equation}
\left | \psi\right >=\frac 1 {\sqrt 2}\left ( \left | 1_{-l}\right >_{\rm A} \left | 1_{l}\right >_{\rm B}+\left | 1_{l}\right >_{\rm A} \left | 1_{-l}\right >_{\rm B} \right ),
\end{equation}
where $l$ is the particle OAM, and the subscripts $\rm A$ and $\rm B$ mean that the first particle is stored by the observer Alice, while the second one is stored by Bob. Then, let Bob send the particle he stores to a Rindler observer moving in the right region, so this particle may be expressed by $ a^{(+)\dagger}_m \left | 0\right >_{\rm B}$ with $m=\pm l$. We require that all modes in this system can be detected with the same energy $\mathcal E>0$. Since $\Omega=(\mathcal E\pm l\omega)/a>0$, the angular velocity of the observer is restricted by $-\mathcal E/l<\omega<\mathcal E/l$. By using the expression for the Minkowski one-particle state in the Rindler coordinates, we can write the entangled state as\begin{align}
\left | \psi \right >=&C \sum_{n_{-l}}\sum_{n_{l}}(-1)^{-ln_{-l}+ln_l}e^{-\pi[(\mathcal E-l\omega)n_{-l}+(\mathcal E+l\omega)n_l]/a}\nonumber \\&\times \bigg [e^{-(\mathcal E+l\omega)\pi/(2a)} \sqrt{(n_l+1)\sinh\left ( \frac {(\mathcal E+l\omega)\pi}{a}\right )}\nonumber \\&\times\left |1_{-l}\right >_{\rm A}\left |n_{-l},n_{l}+1\right >_{\rm R}\left |n_{-l},n_{l}\right >_{\rm L}    +e^{-(\mathcal E-l\omega)\pi/(2a)}\nonumber \\&\times \sqrt{(n_{-l}+1)\sinh\left ( \frac {(\mathcal E-l\omega)\pi}{a}\right )}\left |1_{l}\right >_{\rm A}\left |n_{-l}+1,n_{l}\right >_{\rm R}\nonumber \\&\times\left |n_{-l},n_{l}\right >_{\rm L}\bigg],
\end{align}
where $C=2e^{-2\mathcal E\pi/a}[\coth((\mathcal E-l\omega)\pi/a)-1]^{-1/2}[\coth((\mathcal E+l\omega)\pi/a)-1]^{-1/2}$ is the normalization constant. The density operator may be derived by $\rho=\left | \psi\right >\left < \psi \right |$. Since the entanglement will be measured between Alice and the Rindler observer, we can partially trace out the L part in the density operator, resulting in $\rho^{\rm{AR}}$. Therefore, the system purity, $\mathcal P$, is calculated by \cite{chatterjee2021density,zurek1993coherent,isar1999purity}\begin{align}
\mathcal P=&{\rm{tr}}\big ( (\rho^{\rm{AR}})^2  \big )=\frac 1 4 \left (e^{-2\pi (\mathcal E+l\omega)/a}-1 \right )\Big (-4+e^{-4\mathcal E\pi/a}\nonumber \\& +e^{-4(\mathcal E-l\omega)\pi/a}+3e^{-2(\mathcal E-l\omega)\pi/a}-e^{-2(\mathcal E+l\omega)\pi/a}\Big ).
\end{align}
To measure the entanglement, we first notice that the Rindler particles are thermal and noncorrelated, so we can approximately calculate the entanglement by setting $n_{\pm l}=0$. We will use negativity $\mathcal N$ to quantify the entanglement \cite{guhne2009entanglement,wei2004connections,horodecki2009quantum}. First, we partially transpose the density operator, with respect to the R part, giving\begin{align}
\rho^{\rm{AR}}_{\rm{PT}}=&C^2\big (f_1^2 \left |1_{-l} \right >_{\rm A}\left |1_{l} \right >_{\rm R} \left <1_{-l} \right |_{\rm A}\left <1_{l} \right |_{\rm R}+f_1f_2 \left |1_{-l} \right >_{\rm A}\left |1_{-l} \right >_{\rm R}\nonumber \\&\times \left <1_{l} \right |_{\rm A}\left <1_{l} \right |_{\rm R}+f_1f_2 \left |1_{l} \right >_{\rm A}\left |1_{l} \right >_{\rm R} \left <1_{-l} \right |_{\rm A}\left <1_{-l} \right |_{\rm R}+f_2^2 \nonumber \\&\times\left |1_{l} \right >_{\rm A}\left |1_{-l} \right >_{\rm R} \left <1_{l} \right |_{\rm A}\left <1_{-l} \right |_{\rm R} \big ),
\end{align}
where $f_1=e^{-(\mathcal E+l\omega)\pi/(2a)} \sqrt{\sinh ( (\mathcal E+l\omega)\pi/a)}$ and $f_2=e^{-(\mathcal E-l\omega)\pi/(2a)} \sqrt{\sinh ( (\mathcal E-l\omega)\pi/a)}$. Its eigenvalues are $C^2 f_1^2$, $C^2 f_2^2$, and $\pm C^2 f_1f_2$. Hence, the negativity is given by \begin{align}
\mathcal N=&e^{-5\mathcal E\pi/a}\Big (e^{2(\mathcal E+l\omega)\pi/a}-1 \Big )\Big (e^{2(\mathcal E-l\omega)\pi/a}-1 \Big )\nonumber \\ &\times \sqrt{\sinh \left (\frac { (\mathcal E+l\omega)\pi}{a} \right )\sinh \left (\frac { (\mathcal E-l\omega)\pi}{a} \right )}.
\end{align}
We can do a sense check. When the acceleration $a\rightarrow 0$ and the angular velocity $\omega \rightarrow 0$, the parity $\mathcal P\rightarrow 1$ and the negativity $\mathcal N \rightarrow 1/2$. That means no decoherence or entanglement degradation happens, as one would expect. We plot the purities and the negativities with different OAM $l$ and positive $\omega$ in Fig. \ref{fig.entanglement2D}. One can easily find that there are similar curves for $\omega<0$. The curves show that as the angular velocity $\omega$ increases, the coherence and the entanglement of the system degrade. When $\omega\rightarrow \mathcal E/l$, they will be completely destroyed. From the perspective of the rotating accelerated detector, this is the case where it can interact with an infinite number of thermal particles besides the entangled particle. The detector cannot distinguish thermal and entangled particles; they are totally the same thing. Therefore, the entanglement is completely destroyed. To see the degradation strength for different OAM modes, we define the decay angular velocity as the one by which the negativity of the system drops to $1/e$ of the initial value, i.e., $
\mathcal N(\omega_{\text{D}})=\mathcal N(a=0, \omega=0)/e
$, with a definite acceleration $a$. Numerically, we find that the decay angular velocities are 0.885, 0.442, 0.177, 0.0984, and 0.0443 for $l$=1, 2, 5, 9, and 20, respectively, which shows that higher-order OAM modes will suffer more severe degradation. 

\begin{figure} [tbhp]
\centering
\subfloat[][Purities.]{%
  \includegraphics[width=0.9\linewidth]{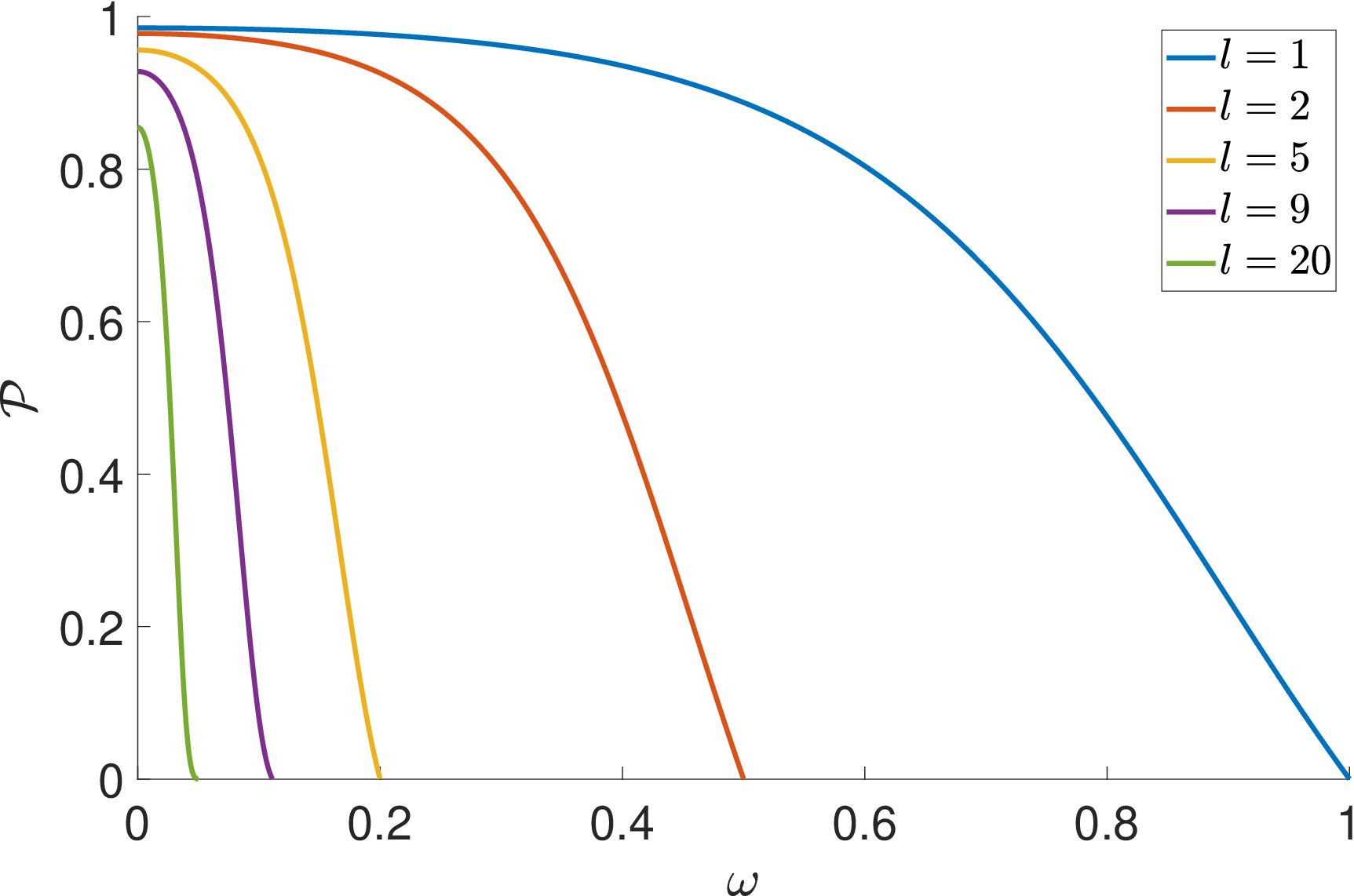}%
  \label{fig.purityHD}
}\hfill
\subfloat[][Negativities.]{%
  \includegraphics[width=0.9\linewidth]{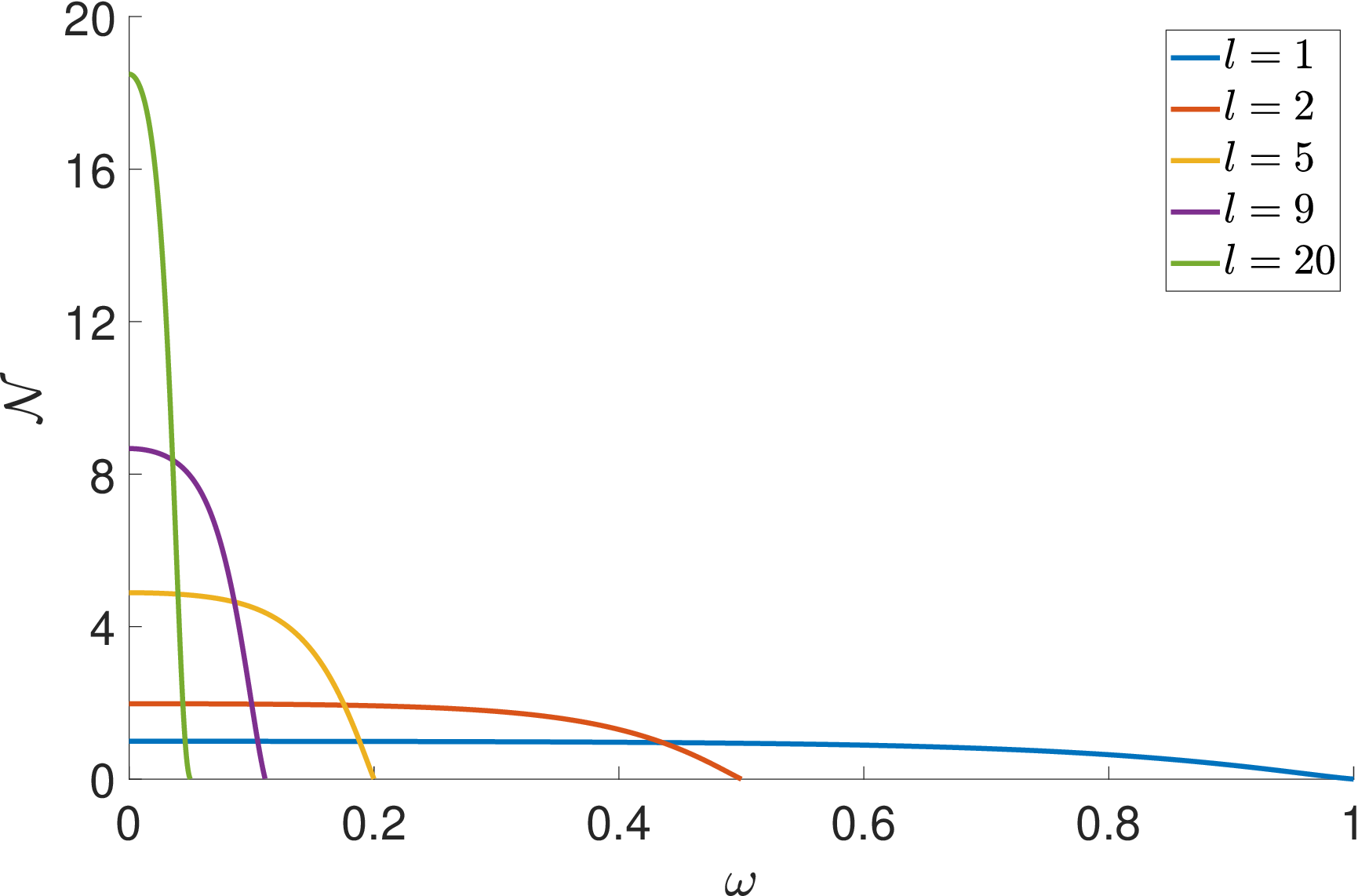}%
  \label{fig.negativityHD}
}\hfill
\caption{Curves for purities and negativities with maximal OAM $M=$1, 2, 5, 9, and 20. We set $\mathcal E=1$ and $a=1$.}
\label{fig.entanglementHD}
\end{figure}

As we have mentioned, the degrees of freedom of OAM have the potential to create high-dimensional entanglement. Will this fact help us further reduce the required $\omega$ if we want to use such entanglement to verify the existence of Rindler particles? Let us study how the high-dimensional entanglement is impacted in the rotating accelerated frame. Suppose we have a high-dimensional two-particle entangled state\begin{equation}
\left | \psi\right >=\frac 1 {\sqrt D}\sum_{l=-M}^M \left | 1_{-l}\right >_{\rm A} \left | 1_{l}\right >_{\rm B} ,
\end{equation}
where $D=2M+1$ is the dimension of entanglement, and again, the subscripts $\rm A$ and $\rm B$ mean the observers Alice and Bob, respectively. Then Bob sends his particle to a Rindler observer moving in the right region. As in the two-dimensional entanglement, similar calculations will give us that the entangled state can be written as\begin{align}
&\left | \psi \right >\nonumber \\ =&   \frac C {\sqrt{D}} \sum_{l=-M}^M  e^{-(\mathcal E+l\omega)\pi/(2a)}\sqrt{2\sinh{\frac {(\mathcal E+l\omega)\pi}{a}}} \sum_{\{n_m=0\}}^\infty   \nonumber \\ &\times(-1)^{\sum_j jn_j} e^{-\pi\sum_j (\mathcal E +j\omega)n_j/a}\sqrt{n_l+1}    \left | 1_{-l}\right >_{\rm A}\nonumber \\ &\times \left |n_{-M},\cdots,n_l+1,\cdots,n_M \right >_{\rm R} \left |n_{-M},\cdots,n_l,\cdots,n_M \right >_{\rm L},
\end{align}
where the index $m$ runs from $-M$ to $M$, the normalization constant $C=\left (\prod_{m=-M}^M \left [ 1-e^{-2\pi (\mathcal E+m\omega)/a}\right ] \right )^{1/2}$, and we only keep track of OAM modes from $-M$ to $M$. The system purity is given by \begin{align}
\mathcal P=&\frac 1 {(2M+1)^2}\prod_{m=-M}^M \tanh \left [\frac {(\mathcal E+m\omega)\pi}{a} \right ] \sum_{l=-M}^M \sum_{j=-M}^M \nonumber \\ &\times\left [ 1+e^{-2(\mathcal E+l\omega)\pi/a} \right ]^{-1} \left [1+e^{-2(\mathcal E+j\omega)\pi/a}   \right ]^{-1},
\end{align}
and the negativity is given by \begin{align}
\mathcal N=&\sum_{l=-M}^M \sum_{j=-M}^{l-1}\frac {2C^2} D \sqrt{ \sinh\left [ {\frac {(\mathcal E+l\omega)\pi}{a}} \right ] \sinh\left [ {\frac {(\mathcal E+j\omega)\pi}{a}}\right ]   } \nonumber \\& \times e^{-(2\mathcal E+l\omega+j\omega)\pi/(2a)}.
\end{align}
The purities and negativities for $M=$1, 2, 5, 9, and 20 are shown in Fig. \ref{fig.entanglementHD}. 

One can find from Fig. \ref{fig.purityHD} that a higher-dimensional system will be impacted more severely when the angular velocity $\omega$ increases. One may notice that, unlike in Fig. \ref{fig.purity2D}, for high-dimensional systems, the purities at $\omega=0$ are different. This reflects the fact that a higher-dimensional system will suffer more degradation when the acceleration $a$ picks up. We will come back to this point soon. One may feel that Fig. \ref{fig.negativityHD} does not clearly show which entanglement degrades faster. So, to quantify the degradation, we can calculate the decay angular velocity $\omega_{\rm D}$. The results are 0.899, 0.449, 0.176, 0.0931, and 0.0368 for $M$=1, 2, 5, 9, and 20, respectively. Compared with the two-dimensional case, higher-dimensional entanglement does not show obvious advantages. It may indicate that the system dimension is insensitive to angular velocity; only the highest order of OAM modes makes a difference. Meanwhile, our previous study \cite{wu2023orbital} shows that for a linearly accelerated observer, if the entanglement dimension is given, then the highest order of OAM modes does not lead to different decay acceleration. In summary, if we want to obtain stronger entanglement degradation with a smaller acceleration, we should increase the entanglement dimension; on the other hand, we should increase the highest order of OAM modes if we want to obtain stronger entanglement degradation with a smaller angular velocity.

\section{Other rotating accelerated frames}

As stated in the Introduction, all stationary trajectories in Minkowski spacetime can be generated by the generators of the Poincar\'e group. Generally, we can write the corresponding Killing field as \begin{equation}
\partial_\tau =\mathbf {P_t}+a\mathbf{K_z}  +\omega_3 \mathbf {J_z}+\omega_1 \mathbf {J_x} \label{eq.general KillingField}
\end{equation}
by choosing appropriate coordinates. Along this trajectory, the rotation direction of the observer may be parallel ($\omega_1=0,~\omega_3\ne 0$), nonparallel ($\omega_1\ne 0$), or orthogonal ($\omega_1\ne 0,~\omega_3= 0$) to the boost direction. 

Actually, we can split the rotating accelerated motion into two cases. The first case is the so-called planar motion with $\omega_1\ne 0,~\omega_3= 0$. This case is rather complicated and should be analyzed by considering two different conditions: $a<\omega_1$ or $a>\omega_1$ \cite{korsbakken2004fulling}. In the former condition, $a<\omega_1$, the motion can be boosted into a pure circular motion, and it has the Minkowski vacuum state. Therefore, there is no OAM spectral for the vacuum state with this motion. Meanwhile, if $a>\omega_1$, the motion can be transformed into a linear accelerated motion along the $z$ direction with a drift with constant velocity along the $y$ direction (Note that our choice of acceleration and rotation directions are different from those in Ref. \cite{korsbakken2004fulling}). Hence, the OAM spectral has a trivial and even distribution, as studied in Ref. \cite{wu2023orbital}.

The other case is the nonplanar motion with $\omega_3\ne 0$. We now show that all nonplanar trajectories can be transformed into the same form. Let us transform the Killing field, Eq. \eqref{eq.general KillingField}, by a boost of $\beta \mathbf {\hat y}$ and a spatial translation of $x_0\mathbf {\hat x}+y_0\mathbf {\hat y}+z_0\mathbf {\hat z}$ (where the hats mean unit vectors) into \begin{align}
\partial_\tau=&-\beta \gamma\omega_3 \mathbf{K_{x'}}+\gamma(a+\beta\omega_1)\mathbf{K_{z'}}+\gamma(a\beta+\omega_1)\mathbf{J_{x'}}\nonumber \\&+\gamma\omega_3\mathbf{J_{z'}}+\gamma(1+\beta\omega_3x_0-az_0-\beta\omega_1z_0)\mathbf{P_{t'}}\nonumber \\&+\omega_3y_0\mathbf{P_{x'}}+[-\gamma\omega_3x_0+\gamma \omega_1z_0+\beta\gamma(az_0-1)]\mathbf{P_{y'}}\nonumber \\&-\omega_1y_0\mathbf{P_{z'}},
\end{align}
with $\gamma=(1-\beta^2 )^{-1/2}$. Now, we require the rotation direction to be parallel to the acceleration direction and all temporal and spatial translation to vanish, i.e., $\gamma\omega_3/[\gamma(a\beta+\omega_1)]=\gamma(a+\beta\omega_1)/(-\beta \gamma\omega_3)$, and all terms before $\mathbf{P_{\mu'}}$ are zero. Then, we can derive the spatial translation to be $x_0=\omega_1/(a\omega_3)$, $y_0=0$, $z_0=1/a$. The restriction that $\left | \beta\right |\le 1$ leaves the solution to $\beta$ as \begin{equation}
\beta=\frac {-(\omega^2_1+\omega^2_3+a^2)+\sqrt{(\omega^2_1+\omega^2_3+a^2)^2-4 a^2\omega^2_1}}{2a\omega_1}.
\end{equation}
By varying $\omega_1$, one can verify that $\left | \beta \right |\le 1$. Thus, we transform all nonplanar motion into motion where the rotation and acceleration directions are parallel, and this motion is discussed in all previous sections. Hence, our previous results can be easily generalized into these nonplanar motions.

\section{Conclusion}
In this paper, we mainly explore the OAM spectrum of Rindler particles in the Minkowski vacuum and the OAM entanglement degradation due to these particles in a rotating accelerated frame. First, the spacetime structure experienced by a rotating accelerated observer is studied. The Rindler coordinates and metric are then constructed. The Klein-Gordon equation is quantized in the cylindrical and the Rindler coordinates. The former gives scalar particles in the usual Minkowski spacetime, while the latter shows that the scalar field can also be expressed by another set of particles, called the Rindler particles. To find the relation between these two different kinds of particles, their Bogoliubov transformation is identified. This somewhat messy transformation can be further written by a single-mode transformation, based on which the OAM spectrum of Rindler particles in the Minkowski vacuum is explored. In contrast with the linear accelerated case, here a new kind of particle with negative energy is allowed to exist. We study how the expected particle numbers for different OAM modes change as the angular velocity $\omega$ varies, and under what conditions the Rindler particles can exist. The results show that the ensemble average of the total OAM approaches zero as $\omega\rightarrow 0$, which is consistent with our earlier study \cite{wu2023orbital}. The total OAM also approaches zero when $\omega \rightarrow \infty$, reflecting the fact that the points of spacetime on the transverse plane are ill-defined in this case. To understand how the rotating accelerated observer (detector) actually experiences, the UD detector is studied. Its transition rate and the detailed balance relation are analyzed. The results can be interpreted in the comoving inertial frame and the rest frame, respectively. As seen in a comoving inertial frame, the results manifest a Boltzmann distribution, but with an energy shift, while from the perspective of the rest frame, the particle numbers of positive- and negative-energy modes affect the state population. Then, the OAM entanglement is explored. We find that the entanglement dimension is insensitive to the angular velocity $\omega$; only the highest order of OAM modes matters. Last, we discuss the general rotating accelerated trajectories and show that we can generalize our results to these stationary trajectories.

There are several questions to be explored. First, the linearly accelerated motion can be related to trajectories with constant radii near a Schwarzschild black hole, so can the rotating accelerated observer be extended to a similar case, e.g., near a rotating black hole? Second, the entanglement dimension and the highest order of OAM modes are mainly affected by acceleration and rotation, respectively. What causes this difference, and do they reflect different aspects of the spacetime structure? Third, there are many sources for gravitational waves and fluctuations. They will introduce dynamical structures into the spacetime. How will the particle definition and OAM spectrum be affected? Normally, the amplitudes of the waves and fluctuations are small, so we anticipate that the particle definition may still be usable in approximation, but the energy and OAM spectra may need to be modified to reflect the presence of the waves and fluctuations. These questions motivate us to deepen our study in this area.

\section*{Acknowledgments}
This work is supported by the National Natural Science Foundation of China (12034016, 11922303), the Natural Science Foundation of Fujian Province of China (2021J02002) and for Distinguished Young Scientists (2015J06002), and the program for New Century Excellent Talents in University of China (NCET-13-0495).

\bibliography{myBib}
\end{document}